\documentclass[aps,prl,reprint,amsmath,amssymb,superscriptaddress]{revtex4-2}

\usepackage[separate-uncertainty]{siunitx}
\usepackage{graphicx}
\usepackage{braket}
\usepackage[T1]{fontenc}
\usepackage{lmodern}
\usepackage[utf8]{inputenc}
\usepackage{hyperref}

\hypersetup{
	pdfpagemode=UseOutlines,
	bookmarksnumbered=true,
	pdflang=en,
	hidelinks,
	hypertexnames=false,
}

\begin{document}
\title{A pulsed ion microscope to probe quantum gases}

\author{C. Veit}
\author{N. Zuber}
\affiliation{5. Physikalisches Institut and Center for Integrated Quantum Science and Technology, Universit\"at Stuttgart, Pfaffenwaldring 57, 70569 Stuttgart, Germany}
\author{O. A. Herrera-Sancho}
\affiliation{5. Physikalisches Institut and Center for Integrated Quantum Science and Technology, Universit\"at Stuttgart, Pfaffenwaldring 57, 70569 Stuttgart, Germany}
\affiliation{Escuela de F\'isica, Universidad de Costa Rica, 2060 San Pedro, San Jos\'e, Costa Rica}
\affiliation{Centro de Investigaci\'on en Ciencia e Ingenier\'ia de Materiales, Universidad de Costa Rica, 2060 San Pedro, San Jos\'e, Costa Rica}
\affiliation{Centro de Investigaci\'on en Ciencias At\'omicas, Nucleares y Moleculares, Universidad de Costa Rica, San Jos\'e, Costa Rica}
\author{V. S. V. Anasuri}
\author{T. Schmid}
\author{F. Meinert}
\author{R. L\"ow}
\author{T. Pfau}
\affiliation{5. Physikalisches Institut and Center for Integrated Quantum Science and Technology, Universit\"at Stuttgart, Pfaffenwaldring 57, 70569 Stuttgart, Germany}
\date{\today}

\begin{abstract}
The advent of the quantum gas microscope allowed for the in situ probing of ultracold gaseous matter on an unprecedented level of spatial resolution. The study of phenomena on ever smaller length scales as well as the probing of three-dimensional systems is, however, fundamentally limited by the wavelength of the imaging light, for all techniques based on linear optics. Here we report on a high-resolution ion microscope as a versatile and powerful experimental tool to investigate quantum gases. The instrument clearly resolves atoms in an optical lattice with a spacing of \SI{532}{nm} over a field of view of 50 sites and offers an extremely large depth of field on the order of at least \SI{70}{\micro m}. With a simple model, we extract an upper limit for the achievable resolution of approximately \SI{200}{nm} from our data. We demonstrate a pulsed operation mode which in the future will enable 3D imaging and allow for the study of ionic impurities and Rydberg physics. 
\end{abstract}

\maketitle

The ability to observe natural phenomena on the single particle level has led to major breakthroughs in modern physics. Prominent milestones in this context were the pioneering experiments in cloud chambers \cite{Gupta46}, the discovery of Rutherford scattering \cite{Geiger13, Rutherford11}, the first imaging of a solid surface with atomic resolution \cite{Muller56} and the observation of quantum jumps for single trapped ions \cite{Bergquist86, Sauter86}. In the realm of quantum gases, time and position sensitive single atom detection enabled Hanbury Brown-Twiss type experiments for both bosons and fermions \cite{Schellekens05, Jeltes07} and a scanning electron microscope was used for the first high-resolution in situ detection of individual atoms \cite{Gericke08, Ott16}. Nowadays, optical quantum gas microscopes offer the possibility to image strongly interacting atoms loaded into optical lattices with single-site resolution \cite{Bakr09, Gross17} and have been used to investigate Bose- and Fermi-Hubbard physics \cite{Bakr10, Sherson10, Mazurenko17, Nichols19}. More recently, ensemble-averaged imaging with super-resolution was demonstrated \cite{McDonald19, Subhankar19}. All microscopes based on linear optics, however, are fundamentally limited in their resolution and depth of field by the wavelength of the imaging light. A high-resolution imaging method capable of extracting three-dimensional information over an extended volume is therefore highly desirable and would allow for the study of quantum correlations, impurity physics and transport phenomena in unprecedented ways.

\begin{figure}[t!]
\centering
\includegraphics[width=\columnwidth]{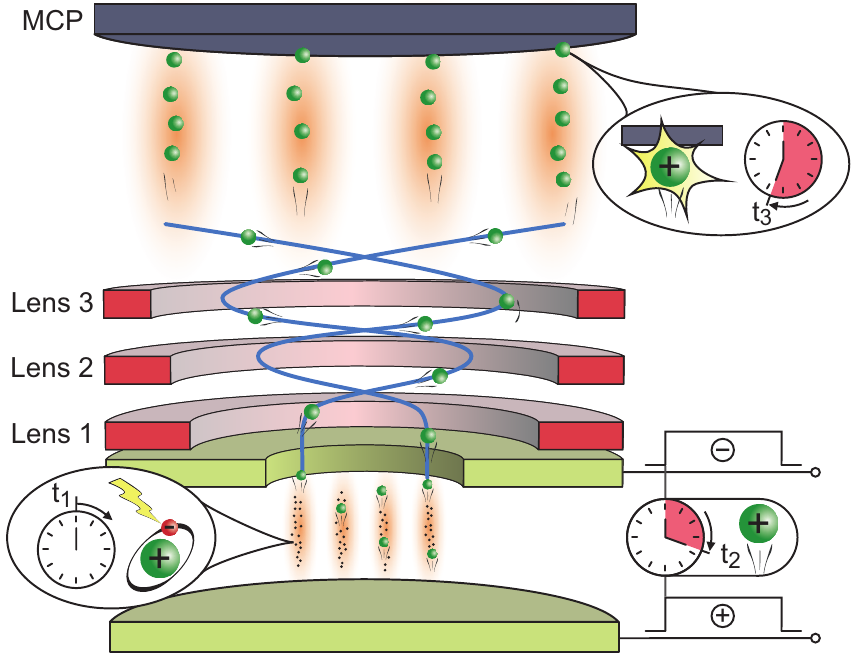}
\caption{Concept of the ion microscope. Neutral atoms in a quantum gas (shown in a 1D periodic potential) are converted into ions at time $t=t_1$ via a suitable ionization process (e.g. near-threshold photo-ionization). After a variable and optional wait time, an extraction field is pulsed on ($t=t_2$) and the ions are imaged by means of an ion microscope consisting of three electro-static lenses onto a spatially resolving detector ($t=t_3$). The high time resolution of the multi-channel plate detector (MCP) allows for the study of dynamical processes (e.g. in strongly interacting many-body systems) and 3D imaging via the time-of-flight information. The versatile nature of the concept enables the study of ground-state ensembles, Rydberg systems and ionic impurities.}
\label{fig:concept}%
\end{figure}

Here we present a high-resolution imaging system based on ion optics, and offering a multitude of intriguing features. Charged-particle optics is commonly employed in cold atom experiments in the form of momentum-space spectrometers \cite{Nguyen04, DePaola08}, but has also been used to image neutral ground-state atoms \cite{Gericke08, Stecker17, Bijnen15} and Rydberg atoms \cite{Schwarzkopf11, Schwarzkopf13, Bijnen15, Fahey15, stecker19}. Our instrument is especially inspired by the ion microscope presented in Ref. \cite{Stecker17}, for which a spatial resolution of smaller than \SI{2.7}{\micro m} was demonstrated, and is based on the following simple concept (Fig. \ref{fig:concept}). Neutral atoms are converted into ions and are subsequently imaged by means of a magnification system onto a spatially and temporally resolving detector. This allows for the investigation of ground-state ensembles, Rydberg excitations and ionic impurities with the very same apparatus. Owing to the excellent time resolution of commercially available ion detectors \cite{Wiza79, Jagutzki02}, both time-dependent measurements and 3D imaging via the time-of-flight information can be realized \cite{Schmid19}. The choice of the ionization process constitutes a powerful degree of freedom, enabling the combination of the high spatial resolution with spectroscopic techniques and permitting weak measurement schemes. Furthermore, near-threshold photo-ionization offers the possibility to produce ultracold ions \cite{Engel18} and opens the door to the spatially resolved study of ionic impurities and ion-atom scattering in the quantum regime \cite{Tomza19, dieterle20, astrakharchik20, feldker20, Schmid18}. In this context, an exciting prospect is the observation of polaron formation and transport dynamics from the two-body collisional timescale to the few- and many-body timescales. We foresee further applications of our approach in the imaging of spatial ordering in three-dimensional Rydberg-blockaded ensembles \cite{Schauss12}, the detection of fine spatial structures in bulk Fermi gases (e.g. Friedel oscillations \cite{Riechers17} and spatial correlations) and the probing of dynamic many-body processes with a high time resolution, just to name a few examples. In the following, we briefly describe our ion microscope and then focus on its performance and imaging characteristics. We begin by discussing a continuous operation mode, in which the object plane is permanently immersed in an extraction field, before demonstrating a pulsed extraction scheme especially suited for the study of ions and Rydberg atoms.
\section{The ion microscope}
\begin{figure}[t!]
\centering
\includegraphics[width=\columnwidth]{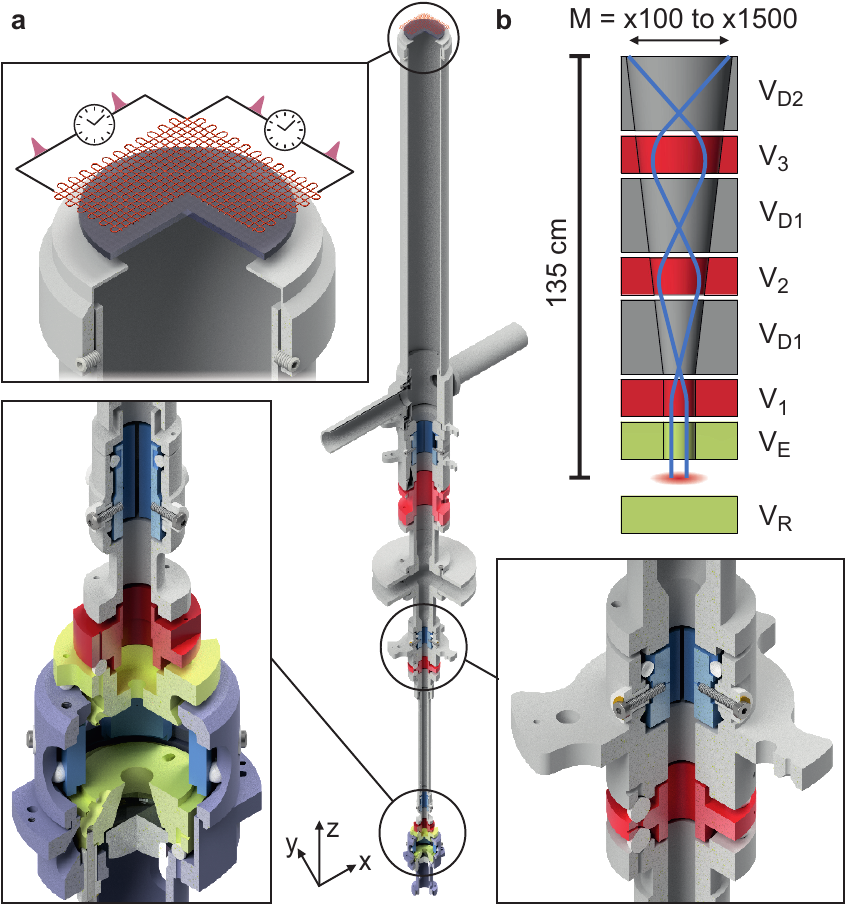}
\caption{Ion microscope. (a) The microscope column consists of three electro-static lenses (main electrodes marked in red, drift sections in gray) and a spatially resolving delay-line detector on top of an MCP stack. The object plane lies between a repeller electrode (hosting an ITO-coated aspheric optical lens) and an extractor electrode which at the same time are part of a six plate electric field control and render the first lens an immersion lens. Each lens is followed by an electro-static quadrupole deflector (blue). (b) Schematic representation of the electrode configuration. Imaging properties and magnification are determined by the extraction voltages ($V_\text{E}$ and $V_\text{R}$), the lens voltages ($V_1$ to $V_3$) and the drift tube voltage ($V_\text{D1}$ and $V_\text{D2}$). The latter are equal for all measurements presented and referred to as $V_\text{D}$. }
\label{fig:microscope}%
\end{figure}
The design of our ion microscope is depicted in Fig. \ref{fig:microscope}a. The whole microscope column has a length of \SI{135}{cm} and consists of three electro-static three-cylinder lenses \cite{Szilagyi12} connected via field-free drift sections. In order to compensate for mechanical tolerances, each lens is followed by a quadrupole deflector. The object plane lies between a repeller and an extractor electrode, the latter of which also acts as the lower lens electrode of the first lens. To allow for the study of Rydberg atoms \cite{Loew12} and ions, the extraction field can be pulsed on a time scale of \SI{30}{ns}. The observation of free ions serves as an excellent probe for stray electric fields, which we cancel by applying small compensation voltages to extractor and repeller as well as to four radially arranged field plates (see Methods). A schematic of the electrode configuration is shown in Fig. \ref{fig:microscope}b. The ions are detected with a multi-channel plate assembly (MCP) in combination with a delay-line detector \cite{Jagutzki02}, offering a high spatial and temporal resolution on the order of \SI{100}{\micro m} and \SI{200}{\pico s}, respectively \cite{Detector}. We expect the detection efficiency of our MCP to be limited to a value close to the open-area ratio of $\sim\!\SI{70}{\percent}$ \cite{Oberheide97} and plan to improve this efficiency in the future by employing a funnel-type MCP offering an open-area ratio of \SI{90}{\percent} \cite{Fehre18}. As detailed in the Methods section, we aim for a further enhancement of the detection efficiency by exploiting facilitated Rydberg excitation around ionic seeds (see Ref. \cite{Urvoy15} and references within). More information on our instrument can be found in Ref. \cite{Schmid19}.

\begin{figure*}[t!]
\centering
\includegraphics[width=\textwidth]{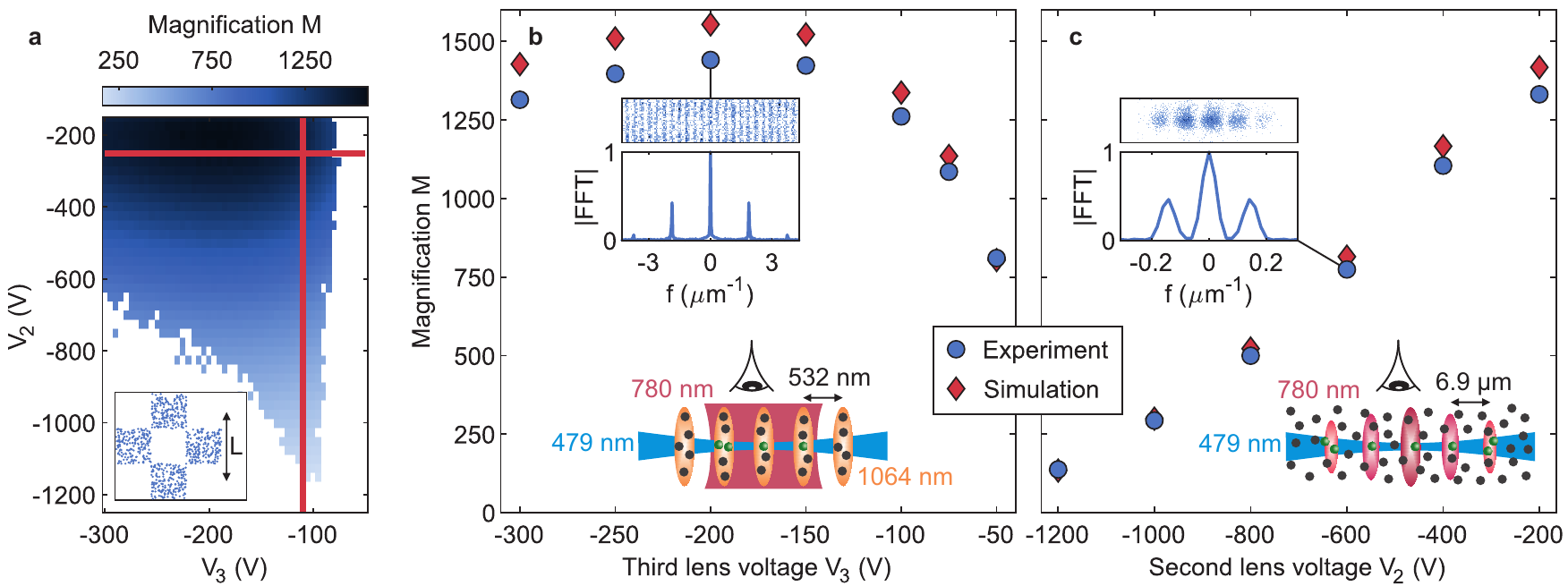}
\caption{Tuning of the magnification. A large range of $M$ is achievable by tuning $V_2$ and $V_3$ ($V_\text{E}=-V_\text{R}=-\SI{500}{V}$, $V_1=\SI{-3.2}{kV}$, $V_\text{D}=\SI{-2.4}{kV}$). (a) Result of numerical trajectory simulations for a test pattern with a line spacing of $L=\SI{100}{nm}$ (see inset) and parameters resembling the experimental situation.  Shown are only data points for which the line spacing is clearly resolved and low imaging distortion is present (see Methods for details). Most values of $M$ are accessible by tuning only $V_2$. (b, c) Comparison between experiment and simulation along the red lines ($V_2=\SI{-250}{V}$, $V_3=\SI{-110}{V}$) shown in (a). (b) The atoms were loaded into a retro-reflected \SI{1064}{nm} lattice and photo-ionized in a crossed beam configuration ($w_{780}\approx\SI{26}{\micro m}$, $w_{479}\approx\SI{5}{\micro m}$). (c) The \SI{780}{nm} photo-ionization laser was spatially structured by projecting the diffraction pattern of a double slit onto the atoms (see Methods). As an example, the insets in (b) and (c) show a cutout of the post-processed detector image along with the normalized magnitude of the fast Fourier transform (FFT) for one of the voltage configurations. The FFT corresponds to the integrated lattice profile of the whole image in object coordinates. $M$ was inferred from the period of the detected patterns. All experimental error bars (see Methods) are significantly smaller than the marker size.}
\label{fig:mag}
\end{figure*}

In contrast to most conventional optical systems, electro-static lens systems offer easy tunablility of imaging properties via the electrode voltages. We explore the parameter space of our ion microscope by means of numerical trajectory simulations performed with a commercially available program \cite{Simion} and find that our system allows for both 2D and 3D imaging \cite{Schmid19}. In a first step, we focus here on the performance of our microscope regarding 2D imaging. In this mode, we keep all drift tubes at the potential of the MCP front plate (\SI{-2.4}{kV}) and employ typical extraction voltages of $V_\text{E} = -V_\text{R} = \SI{-500}{V}$. Owing to their small initial velocity, the cold ions are extracted on trajectories almost parallel to the extraction field. Consequently, a sharp projection of the object plane can be observed in the detector plane even for voltage configurations for which no actual image formation occurs. 
%
\section{Characteristics of the imaging system}
In the following, we present experimental results on the characteristics of the previously described imaging system. For most of the measurements, $^{87}$Rb atoms were loaded into a one-dimensional optical lattice with a spacing of \SI{532}{nm} and photo-ionized in the crossing volume of two laser beams with wavelengths of \SI{780}{nm} and \SI{479}{nm}. The \SI{780}{nm} beam was blue-detuned by $\SI{78}{MHz}$ from the $\ket{5P_{3/2}, F=3}$ intermediate state and aligned parallel to the optical axis of the microscope (z-direction, see Fig. \ref{fig:microscope}a). The \SI{479}{nm} beam was aligned perpendicular to the optical axis (y-direction) and tuned such as to realize a near-threshold ionization with typical excess energies on the order of $h\times \SI{100}{GHz}$ or below, where $h$ corresponds to the Planck constant. The above described setup was used to characterize the magnification, field of view (FOV), depth of field (DOF) and resolution of our microscope. The waists of the ionization beams were shaped to suit the specific measurement. For magnifications too small to resolve the optical lattice, we held the atoms in an optical dipole trap and induced a spatial structure of the \SI{780}{nm} light field by projecting the diffraction pattern of a double slit onto the atoms. The period of the resulting ionization pattern was measured at large magnifications using the optical lattice as a ruler. All measurements presented are integrated over several experimental cycles and, as detailed later, compensated for phase drifts of the optical lattice as well as for minor distortions of the detected images. Additional information on the experiment and the data evaluation are given in the Methods section.
%
\subsection{Magnification}
To explore the exceptional tunability of the magnification $M$ of our instrument, we fixed all electrode voltages but $V_2$ and $V_3$ (see caption of Fig. \ref{fig:mag}) and used the latter two to tune the focal lengths of the second and third lens. Figure \ref{fig:mag}a shows a corresponding simulation of the total magnification revealing that high-quality imaging results can be obtained for magnifications ranging from below $\times 200$ to above $\times 1500$. Both higher and lower magnifications can be achieved if additional electrode voltages are adjusted. We compare the behavior of our microscope with the simulation for scans of either $V_2$ (Fig. \ref{fig:mag}c) or $V_3$ (Fig. \ref{fig:mag}b) for which the respective other voltage was fixed (see red lines in Fig. \ref{fig:mag}a). As delineated above, the magnification was extracted by imaging either atoms in an optical lattice, or a diffraction pattern of known period (see insets). Evidently, the simulations describe our imaging system fairly accurate with the largest deviation from the experiment being smaller than \SI{10}{\percent}. For our detector diameter (\SI{40}{mm}), the demonstrated magnification range maps to a FOV between approximately \SI{30}{\micro m} and \SI{300}{\micro m}. This permits both high-resolution studies and the observation of large atomic ensembles.
%
\subsection{Field of view}
\begin{figure}[t!]
\centering
\includegraphics[width=\columnwidth]{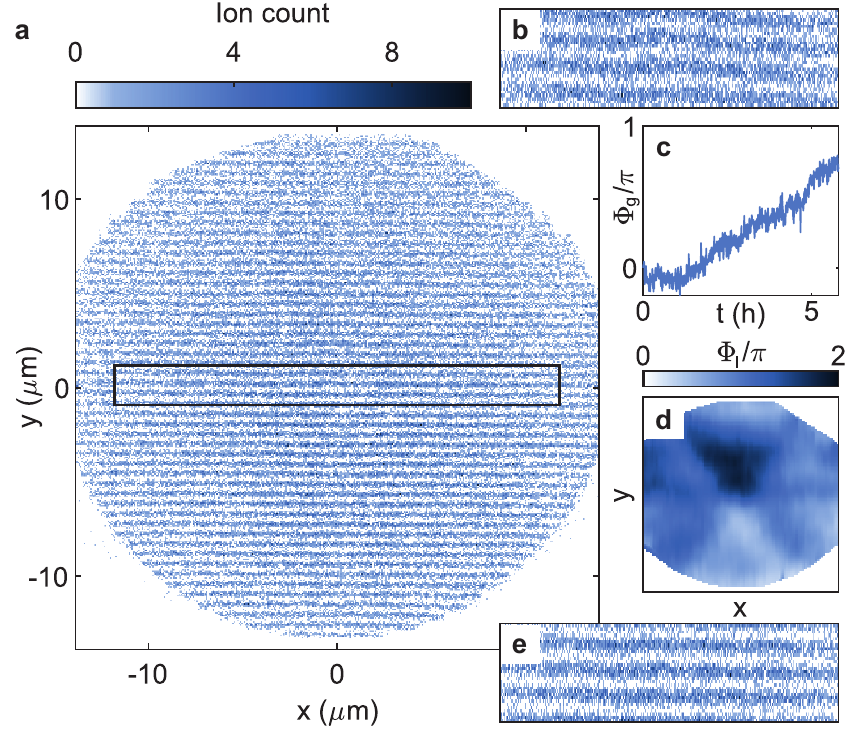}
\caption{Field of view. The FOV was investigated for the voltage configuration corresponding to the largest magnification shown in Fig. \ref{fig:mag}b ($M=1441$). (a) Measured data (sum over 1200 experimental cycles, object coordinates), corrected for a global phase $\Phi_\text{g}$ (drift of the lattice) and a local phase $\Phi_\text{l}$ (distortion). The marked region is shown enlarged in (b) (raw data) and (e) (phase-corrected data). (c) $\Phi_\text{g}$ over time. (d) $\Phi_\text{l}$ is a function of the spatial coordinates and was extracted from the local deviation of the experimental data from the expected regular lattice structure.}
\label{fig:fov}
\end{figure}
In order to confirm that the large magnifications suitable for high-resolution imaging can be made use of without any sacrifices concerning the FOV, we utilized our optical lattice as a test pattern. To this end, we employed a large \SI{780}{nm} beam ($w_{780} \approx \SI{36}{\micro m}$) and shaped the \SI{479}{nm} beam to a horizontal light sheet with a waist of $w_{x,479} \approx \SI{40}{\micro m}$. The measurement shown in Fig. \ref{fig:fov}a illustrates that the lattice can be clearly resolved over the whole detector area, corresponding to a FOV of 50 lattice sites. The data shown is post-processed in two ways (see also Methods). First, we make use of the regular structure of the lattice to compensate for a local phase $\Phi_\text{l}$ caused by small distortions common to all measurements (Fig. \ref{fig:fov}d). Second, we compensate for a time dependent global phase $\Phi_\text{g}$ caused by a thermal drift of the lattice over the measurement time of almost six hours (Fig. \ref{fig:fov}c). We find that the local phase $\Phi_\text{l}$ stays below $2\pi$ over the whole detector and use the data shown in Fig. \ref{fig:fov}d to compensate for the observed distortion in all our measurements. A comparison between the raw data and the post-processed data is shown in Figs. \ref{fig:fov}b,e for the marked region in Fig. \ref{fig:fov}a. Even the raw data shows a good contrast.
%
\subsection{Depth of field}
Field of view and DOF of an imaging system determine the maximum dimensions of the objects that can be imaged. In the case of high-resolution optical microscopes, the imaging volume is typically restricted by a small DOF originating from the large numerical aperture of the systems. In contrast, the DOF of our ion microscope is remarkably large as demonstrated in the following. The DOF was probed for the same voltage configuration as used for the FOV measurement by employing a large \SI{780}{nm} beam ($w_{780} \approx \SI{36}{\micro m}$) and by shaping the \SI{479}{nm} beam to a vertical light sheet ($w_{x, 479} \approx \SI{2}{\micro m}$, $w_{z, 479} \approx \SI{100}{\micro m}$). Due to the large extent of the light sheet in z-direction, the height of the ionization region was primarily limited by the diameter of the atomic cloud ($1/e$-diameter of approximately \SI{70}{\micro m}). Careful alignment of the lattice planes was required in order to attain a good imaging contrast. A resulting measurement is shown in Fig. \ref{fig:dof_fft}a along with the normalized magnitude of the FFT corresponding to the integrated lattice profile (Fig. \ref{fig:dof_fft}b). As expected from simulations, we find that the imaging result does not suffer from the increased DOF and is very comparable to the results obtained from the magnification measurements (see inset of Fig. \ref{fig:mag}b for comparison). Given the diameter of the atomic cloud, we can quantify the DOF to be on the order of at least \SI{70}{\micro m}. In comparison with quantum gas microscopes, typically featuring a DOF smaller than a few micrometers, the DOF or our ion optics is exceptionally large and allows for the study of three-dimensional bulk gases. 
%
\subsection{Resolution}
\begin{figure}[t!]
\centering
\includegraphics[width=\columnwidth]{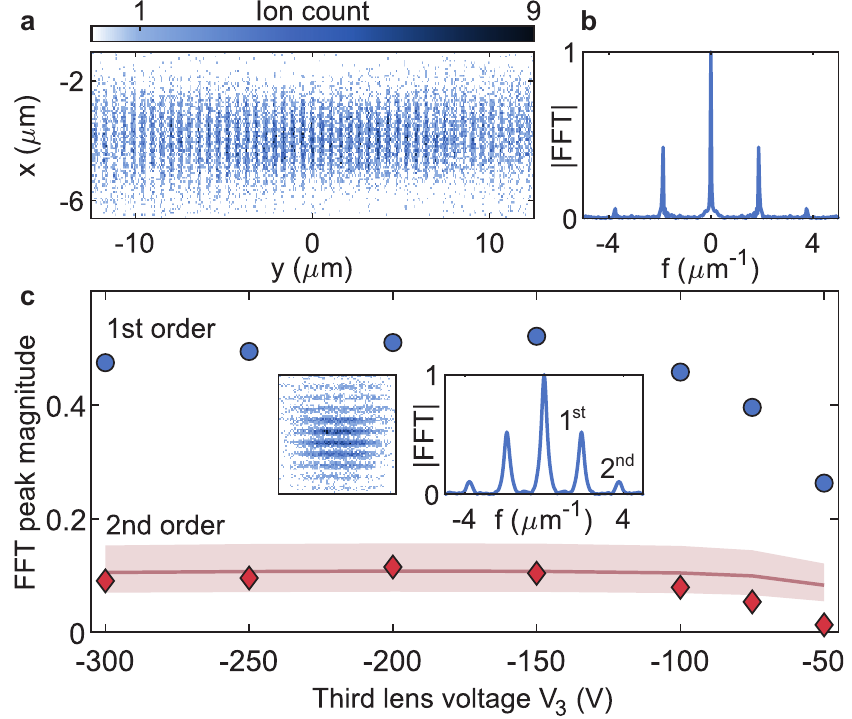}
\caption{Depth of field and resolution. (a) Imaging result for an increased depth of field limited only by the cloud size ($1/e$-diameter of \SI{70}{\micro m}). (b) Absolute value of the FFT corresponding to the integrated lattice profile shown in (a). (c) The magnitudes of the first and second-order FFT peaks (here as a function of $V_3$ and for a tightly confined ionization region) serve as a qualitative measure of the resolution. All experimental error bars (see Methods) are smaller than the marker size. The insets show the experimental data for $V_3 = \SI{-150}{V}$. Solid line and shaded area: theoretical prediction for the upper bound of the second-order peak amplitude for a resolution of $200\pm20\,\text{nm}$. Experimental circumstances are detailed in the main text.}
\label{fig:dof_fft}%
\end{figure}
From the measurements presented above it is evident that, for large magnifications, the resolution of the microscope is significantly smaller than the period of our test target given by the optical lattice. In order to still characterize the resolution, we take advantage of the localization of the atoms caused by the tight confinement in the optical lattice. The signature of this localization is found in the FFT of our measured data, in which a signal at twice the lattice frequency is apparent (see e.g. Fig. \ref{fig:dof_fft}b). We use the amplitude of this second-order peak as a qualitative measure of the resolution $r$ of our microscope and use a simple model to give a quantitative upper limit of $r$. Due to the finite detector resolution, we expect our imaging system to perform best at large magnifications. Consequently, the voltage configuration corresponding to the data shown in Fig. \ref{fig:mag}b was chosen for our resolution studies. In order to minimize effects of residual distortions, a tightly confined photo-ionization volume was used for the measurement ($w_{780}\approx\SI{2}{\micro m}$, $w_{479}\approx\SI{5}{\micro m}$). Figure \ref{fig:dof_fft}c shows the magnitude of the first- and second-order FFT peak corresponding to the measured data as a function of $V_3$. The significant amplitude of the second-order peak indicates an excellent resolution over almost the whole measurement range. The drop of contrast at small magnitudes of $V_3$ is in accordance with our simulations and results from both imaging aberrations and a decrease of magnification. To extract an upper limit of the resolution, we assume all atoms to occupy the lowest band of the optical lattice and employ a Gaussian approximation for the single-site wave function (see Methods for details). We then convolve the density profile with a Gaussian point spread function (PSF), scale it according to the magnification of our microscope and bin the profile with the bin size used for our experimental data (\SI{100}{\micro m}). Expectedly, the FFT of the model profile shows a second-order peak at twice the lattice frequency, enabling comparison with the experiment. We identify the full width at half maximum of the PSF as the resolution $r$ and find that, for the highest magnifications, we get good agreement with our experimental data for $r\approx \SI{200}{nm}$ (see solid line in Fig. \ref{fig:dof_fft}c). In consideration of the non-optimized loading procedure of the optical lattice, the atoms most certainly occupy several bands instead of only the lowest band. Since this leads to a less confined wave function (and therefore to a smaller amplitude of the second-order FFT peak), the actual resolution of our microscope is probably significantly smaller than \SI{200}{nm}. Indeed, numerical simulations accounting for realistic mechanical tolerances and voltage noise suggest that a resolution on the order of \SI{100}{nm} is achievable \cite{Schmid19}. This permits the in situ observation of phenomena taking place at the length scale of the healing length of Bose-Einstein condensates \cite{Abo01} or the Fermi wavelength of ultracold Fermi gases \cite{Riechers17}. 
%
\subsection{Pulsed operation}
\begin{figure}[t!]
\centering
\includegraphics[width=\columnwidth]{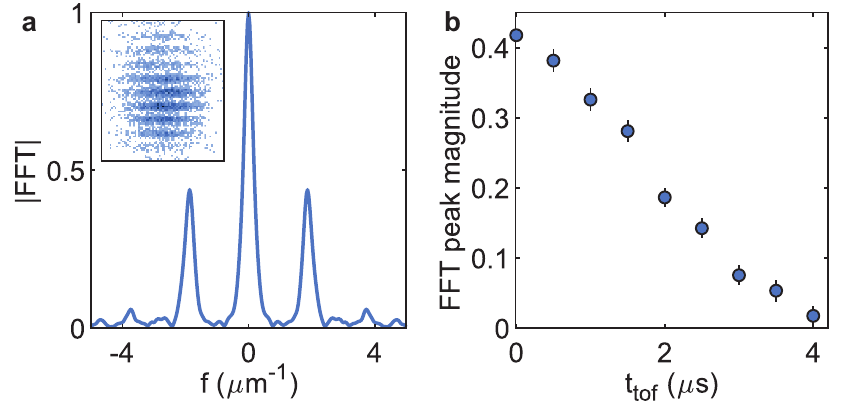}
\caption{Pulsed operation mode. (a) Imaging result for $t_\text{tof}=\SI{0}{\micro s}$ and magnitude of the FFT corresponding to the integrated lattice profile. The clear signature of a second-order Fourier peak indicates a resolution on the order of half the lattice spacing (\SI{266}{nm}). (b) The decay of the first-order FFT peak serves as an indicator for the blurring of the lattice structure as a function of $t_\text{tof}$. The timescale of the decay reveals the cold temperature of the produced ions in the few tens of microkelvin range. Error bars correspond to a conservative estimate of the statistical error (see Methods).}
\label{fig:pulsed}%
\end{figure}
A powerful feature of our ion microscope is the ability to directly image ions and field-ionized Rydberg atoms. Both the study of Rydberg and ion physics, however, would be hindered by a constant extraction field. Therefore, our instrument is designed such as to allow for a pulsed extraction. In the measurements presented in the following, a fast high-voltage switch was employed to toggle between small compensation voltages and large extraction voltages being applied to extractor and repeller (Fig. \ref{fig:microscope}). By the additional use of four radial field plates, this procedure enables us to precisely cancel stray electric fields while the extraction field is switched off. The pulsed operation mode was tested for the same optical configuration as was used for the measurements discussed in the previous paragraph and slightly different voltage settings resulting in a magnification of $M=1467$ (see Methods). In every experiment, 6000 ionization cycles were realized, each consisting of a \SI{1}{\micro s} long photo-ionization pulse followed by a variable wait time $t_\text{tof}$, after which the extraction voltage was switched on. A measurement result for $t_\text{tof}=\SI{0}{\micro s}$ is shown in the inset of Fig. \ref{fig:pulsed}a together with the corresponding FFT. As for the data measured in the continuous operation mode, second-order Fourier peaks are clearly observable and indicate a resolution on the order of half the lattice spacing.
For increasing wait time $t_\text{tof}$, the visibility of the lattice structure decreases and a decay of the first-order Fourier peak can be observed (Fig. \ref{fig:pulsed}b). This approximately Gaussian-shaped decay with a $1/e$-timescale of \SI{2.4}{\micro s} is consistent with an ion temperature of a few tens of microkelvin. We attribute the associated kinetic energy mostly to the excess ionization energy ($\sim \! k_\text{B} \times \SI{13}{\micro K}$, with $k_\text{B}$ being the Boltzmann constant) and the occupancy of several excited bands in the optical lattice. A reduction of the ionization energy in combination with an ultracold atomic sample will in the future allow for the creation of much colder ions and enable the investigation of ion-atom hybrid systems in the quantum regime.
%
\section{Conclusion and outlook}
We have presented a high-resolution ion microscope allowing for the time-resolved probing of quantum gases on a single atom level. The magnification of the imaging system was shown to be highly tunable, enabling the investigation of both isolated microscopic few-body processes and extended many-body systems. With a resolution better than \SI{200}{nm} and an exceptionally large DOF of more than \SI{70}{\micro m}, our microscope is excellently suited to study bosonic and fermionic bulk quantum gases on the length scale of the healing length and the Fermi wavelength, respectively. A pulsed operation mode enables the spatially resolved study of ion-atom hybrid systems and Rydberg ensembles and permits the ultra-precise measurement and subsequent compensation of stray electric fields.

We believe that charged-particle optics holds great promise for the field of ultracold quantum gases and will allow for a whole range of new experimental techniques as well as an unprecedented level of precision. Key aspects in this respect are the exceptional spatial and temporal resolution, the possibility of 3D imaging and the freedom to combine the spatial resolution with spectroscopic techniques (e.g. enabling spin-resolved detection). 

As a next step, we plan to use the 3D imaging capabilities of our apparatus in combination with near-threshold photo-ionization to create and study ultracold ionic impurities in a degenerate quantum gas \cite{Tomza19, astrakharchik20}. Using the same experimental tools, we also aim for the observation of individual ion-atom collisions in the quantum regime \cite{Schmid18}.

\section{Acknowledgments}
We thank H. Rose for a valuable discussion and T. Dieterle, C. Tomschitz and P. Kongkhambut for important contributions in the earlier stages of this work. We are further indebted to our workshops for the manufacturing of the ion microscope and the HV power supplies. We  acknowledge  support  from  the Deutsche Forschungsgemeinschaft [Projects No. PF 381/13-1 and No. PF 381/17-1, the latter being part of the SPP 1929 (GiRyd)] and have received funding from the QuantERA ERANET programme under the project "Theory-Blind Quantum Control - TheBlinQC". C.V. gratefully acknowledges support from the Carl Zeiss Foundation. F.M. is indebted to the Baden-Württemberg-Stiftung for the financial support by the Eliteprogramm for Postdocs and O.A.H.S. and F.M. are indebted to the support by the Carl Zeiss Foundation via IQST.
%
%
%
\appendix
\section{Methods}
\textbf{Trajectory simulations.} The imaging properties of the ion optics were simulated by propagating test patterns through the electro-static potential corresponding to a given electrode voltage configuration. For the simulation shown in Fig. \ref{fig:mag}a, two test patterns with a lateral line spacing of $L=\SI{100}{nm}$ and an extent of \SI{10}{\micro m} along the optical axis were employed (see inset for the shape of the patterns). Each pattern consisted of 800 ions given an isotropic velocity distribution corresponding to an initial kinetic energy of $k_\text{B} \times \SI{10}{\micro K}$. One of the patterns was centered on the optical axis, whereas the other one was offset by a distance corresponding to \SI{42.5}{\percent} of the FOV. Here, the FOV is defined as the detector diameter divided by the magnification. Shown are only data points for which the test patterns are clearly resolved and for which the local magnification extracted from the off-axis pattern differs by less than \SI{3}{\percent} from the on-axis case. The latter condition aims to ensure a low distortion of the imaging result. Since the effects of mechanical asymmetries and voltage noise are not captured by the presented simulation, no quantitative prediction can be made for the resolution.

\textbf{Preparation of the atomic cloud.}
Each experimental cycle started with the loading of approximately \num{4e8} $^{87}$Rb atoms from an effusive oven into a magneto-optical trap (MOT). For this, a double-species Zeeman slower was employed which in the future will allow us to perform experiments with Lithium \cite{Marti10}. The atoms were then optically transported along the x-axis (Fig. \ref{fig:microscope}) from the MOT chamber into a separate science chamber, above which the ion microscope is located. This was accomplished by the use of a transport trap consisting of two \SI{1064}{nm} laser beams passing through a lens mounted on an air-bearing translation stage \cite{Gross16}.  The beams cross under a small angle of \SI{2.3}{\degree}, producing a transverse intensity profile with a $1/e^2$-waist of approximately \SI{70}{\micro m} at the crossing point. For the measurements performed at low magnifications (Fig. \ref{fig:mag}c), we held typically \numrange{2e6}{3e6} atoms at a temperature of approximately \SI{8}{\micro K} in the transport trap. For the experiments performed in an optical lattice, we ramped down the power of the transport trap on a time scale of \SI{1.5}{s} and ramped up the lattice within \SI{100}{ms}. The latter was produced by a retro-reflected \SI{1064}{nm} laser beam oriented along the y-axis (Fig. \ref{fig:microscope}) and possessing a $1/e^2$-waist of $w_{1064} \approx \SI{110}{\micro m}$. After the ramping process, the atoms were solely held by the optical lattice potential with a depth of $\sim\! 1100\, E_\text{r}$, where the recoil energy is defined as $E_\text{r} = \hbar^2 k^2/2m$. Here $m$ corresponds to the atomic mass, $\hbar$ is the reduced Planck constant and $k$ is the wave vector of the lattice light. All experiments were performed in unpolarized samples with the atoms being pumped to the $F=2$ hyperfine state. Depending on the specific measurement, a complete experimental cycle lasted between $10$ and \SI{13}{s}.

\textbf{Photo-ionization sequence.}
After the preparation of the atomic cloud, a photo-ionization phase followed. For the measurements performed in the continuous operation mode of the microscope, the ionization sequence lasted \SI{200}{ms}, during which time the \SI{780}{nm} laser was switched on continuously. Due to technical reasons, the \SI{479}{nm} laser was pulsed with a duty cycle of \SI{95}{\percent} and a frequency of \SI{50}{kHz}.  For the measurements conducted in the pulsed operation mode, the ionization sequence lasted \SI{1.8}{s} and consisted of 6000 ionization cycles. In each cycle, a \SI{1}{\micro s} long photo-ionization pulse was followed by a variable wait time $t_\text{tof}$ (see main text), before the extraction field was pulsed on. In order to minimize the effect of dark counts which occur at a typical rate of a few Hertz, the detector was gated such as to register only signals compatible with the employed ionization sequence. The number of detected ions per experimental cycle (i.e. for one atomic cloud) depended on the specific measurement and ranged typically between a few tens and a few hundreds of ions. As an example, the measurement shown in Fig. \ref{fig:dof_fft}a consists of 18400 ion counts which were acquired over the course of one hour. 

\textbf{Ion detection.}
The employed delay-line detector offers a spatial and temporal resolution on the order of \SI{100}{\micro m} and \SI{200}{ps}, respectively, and features a maximum continuous detection rate of \SI{1}{MHz} with a multi-hit dead time on the order of 10 to \SI{20}{ns} \cite{Detector}. The delay-line is combined with two stacked MCPs with an active diameter slightly larger than \SI{40}{mm} and an open-area ratio (OAR) of \SI{70}{\percent} (F1217-01 MOD7, Hamamatsu). Since the detection efficiency $\eta$ of MCPs is typically limited to values close to the OAR \cite{Oberheide97}, we plan to exchange the front MCP with a funnel-type model offering an OAR of \SI{90}{\percent} \cite{Fehre18}.  In addition, we envision to enhance the overall quantum efficiency to detect a single ion to higher than \SI{99}{\percent} using a Rydberg antiblockade effect (see Ref. \cite{Urvoy15} and references therein) which allows for the resonant laser excitation of Rydberg atoms on a shell around the ion. Due to the Stark effect of Rydberg atoms \cite{Loew12}, this shell can have a radius on the order of \SI{1}{\micro m}. As the extraction field of the ion microscope is pulsed on, the Rydberg atoms are field ionized and the initially single ion is converted into $n$ ions (typically $n=2-6$), such that the infidelity of the detection is reduced to $(1-\eta)^n$. Since the time scale for the Rydberg excitation can be very fast ($\sim\!10$ to $\SI{100}{ns}$), the proposed scheme can be repeated several times in a single atomic cloud. The secondary ions responsible for the described preamplification process share on average a common center of gravity with the initial ion.  However, depending on the amplification protocol, a small position noise as well as a loss of temporal resolution is introduced. For the specific application, a trade-off between spatial resolution, time resolution and quantum efficiency will have to be found.

\textbf{Determination of the magnification.}
For the measurements performed in an optical lattice (see Fig. \ref{fig:mag}b), the FFT corresponding to the integrated lattice profile was fitted assuming a Gaussian shape of the first- and second order peaks evoked by the periodicity of the lattice. The magnification was then determined from the fitted peak positions and the known wavelength of the lattice laser $\lambda_\text{lat}=\SI{1063.93(1)}{nm}$. Error bars were determined by considering the conservative estimate of the systematic error of $\lambda_\text{lat}$ and the standard error of the fitted peak positions. For the characterization of small magnifications for which the lattice could not be resolved (see Fig. \ref{fig:mag}c), a spatial structure was induced into the \SI{780}{nm} photo-ionization beam by passing the beam through a double slit (slit separation and width of \SI{3}{mm} and \SI{1}{mm}, respectively), before focusing the beam onto the atoms. To this end, an in-vacuum aspheric lens with an effective focal length of \SI{26}{mm} was used. The described setup resulted in an image of the far-field diffraction pattern of the double slit in the focal plane of the lens. The period of the diffraction pattern $\lambda_\text{diff}$ was measured at a known magnification previously calibrated with the optical lattice. In comparison with the measurements performed in an optical lattice, the \SI{479}{nm} beam was enlarged in x-direction to $w_{479x}\approx\SI{100}{\micro m}$ and the \SI{780}{nm} beam measured approximately $w_{780x}\approx\SI{2}{\micro m}$ at the position of the atoms. The magnification was extracted by fitting the imaged diffraction pattern. Error bars were determined by considering the error of $\lambda_\text{diff}$ and the standard error of the fitted pattern period. All error bars in Fig. \ref{fig:mag} are significantly smaller than the marker size.

\textbf{Data processing.}
The employed detector delivers a time stamp and a position for each detected ion. We compensated for minor distortions of the detector image in y-direction (Fig. \ref{fig:microscope}) by applying a spatially dependent shift to the detected y-position of all ions. The spatial distribution of this shift was found by using the regular structure of the optical lattice as a test pattern (Fig. \ref{fig:fov}). The local distortion could be interpolated from the discrepancy between the observed lattice phase and the phase expected for an ideal lattice structure (Fig. \ref{fig:fov}d). We have checked that the influence of wave-front aberrations of the lattice beams is small by shifting the image along the x-direction of the detector using the electro-static deflector after the third lens (Fig. \ref{fig:microscope}). To compensate for thermal drifts of the optical lattice, we sorted the detection events into groups of 100 ions and calculated the FFT corresponding to the integrated lattice profile for each of the groups. We then extracted the global phase of the lattice from the complex amplitude of the first-order FFT peak and corrected for the drift accordingly. Subsequently, the data was binned with a bin size of \SI{100}{\micro m}. For measurements for which the magnitudes of the FFT peaks were extracted, we determined a statistical error of these by considering again groups of 100 ions. For each of the groups, the complex amplitudes of the first- and second-order FFT peaks were determined. An estimate for the error of the peak magnitudes was then found by dividing the standard deviation of the amplitudes by $\sqrt{n_\text{g}}$, where $n_\text{g}$ is the number of groups.

\textbf{Pulsed operation mode.} The measurements shown in Fig. \ref{fig:pulsed} were acquired using the following electrode voltages:
$[V_\text{R}$, $V_\text{E}$, $V_1$, $V_2$, $V_3$, $V_\text{D}$] = [$400$, $-400$, $-2800$, $-200$, $-110$, $-2400]\, \si{V}\,.$ For the pulsed measurements shown in Fig. \ref{fig:pulsed}b, the magnitude of the first-order FFT peak decreased with increasing $t_\text{tof}$. In order to still accurately compensate for the thermal drift of the lattice, a measurement for $t_\text{tof}= \SI{0}{\micro s}$ was performed in every second experiment cycle. The thermal drift of the lattice was then deduced only from these additional measurements and corrected for according to the procedure detailed in the previous paragraph.

\textbf{Approximation of the single-site wave function.}
For atoms occupying the lowest energy band of our deep optical lattice, we approximate the single-site wave function by a ground-state harmonic oscillator wave function. To this end, we consider the Taylor expansion of the lattice potential around an energy minimum $V(x) \approx V_0 k^2 x^2$ and find the probability density of the lowest harmonic oscillator state to be 
\begin{equation}
|\psi(x)|^2 = \frac{(2 m V_0 k^2)^{1/4}}{(\pi \hbar)^{1/2}} \exp\left( -\frac{\sqrt{2mV_0 }\,k}{\hbar} x^2\right)\, .
\end{equation}
Here $V_0$ is the depth of the lattice potential, $m$ is the atomic mass and $k$ is the wave vector of the \SI{1064}{nm} lattice light.

\textbf{Compensation of stray electric fields.}
For the presented measurements performed in the pulsed operation mode, stray electric fields within the ionization volume were compensated by applying suitable voltages to the six compensation electrodes. In order to calibrate these voltages on the order of a few tens of millivolts, free ions were observed for up to \SI{70}{\micro s}. The compensation voltages were adjusted such as to minimize the displacement of the detected ion distribution as the observation time was increased. For the field component along the optical axis, the time-of-flight information from the ion detection was employed. From the observed spread of the ion cloud, the typical magnitude of residual fields in the transverse direction can be quantified to be on the order of \SI{100}{\micro V/cm} over a spherical volume with \SI{20}{\micro m} diameter. 

\begin{thebibliography}{47}%
\makeatletter
\providecommand \@ifxundefined [1]{%
 \@ifx{#1\undefined}
}%
\providecommand \@ifnum [1]{%
 \ifnum #1\expandafter \@firstoftwo
 \else \expandafter \@secondoftwo
 \fi
}%
\providecommand \@ifx [1]{%
 \ifx #1\expandafter \@firstoftwo
 \else \expandafter \@secondoftwo
 \fi
}%
\providecommand \natexlab [1]{#1}%
\providecommand \enquote  [1]{``#1''}%
\providecommand \bibnamefont  [1]{#1}%
\providecommand \bibfnamefont [1]{#1}%
\providecommand \citenamefont [1]{#1}%
\providecommand \href@noop [0]{\@secondoftwo}%
\providecommand \href [0]{\begingroup \@sanitize@url \@href}%
\providecommand \@href[1]{\@@startlink{#1}\@@href}%
\providecommand \@@href[1]{\endgroup#1\@@endlink}%
\providecommand \@sanitize@url [0]{\catcode `\\12\catcode `\$12\catcode
  `\&12\catcode `\#12\catcode `\^12\catcode `\_12\catcode `\%12\relax}%
\providecommand \@@startlink[1]{}%
\providecommand \@@endlink[0]{}%
\providecommand \url  [0]{\begingroup\@sanitize@url \@url }%
\providecommand \@url [1]{\endgroup\@href {#1}{\urlprefix }}%
\providecommand \urlprefix  [0]{URL }%
\providecommand \Eprint [0]{\href }%
\providecommand \doibase [0]{https://doi.org/}%
\providecommand \selectlanguage [0]{\@gobble}%
\providecommand \bibinfo  [0]{\@secondoftwo}%
\providecommand \bibfield  [0]{\@secondoftwo}%
\providecommand \translation [1]{[#1]}%
\providecommand \BibitemOpen [0]{}%
\providecommand \bibitemStop [0]{}%
\providecommand \bibitemNoStop [0]{.\EOS\space}%
\providecommand \EOS [0]{\spacefactor3000\relax}%
\providecommand \BibitemShut  [1]{\csname bibitem#1\endcsname}%
\let\auto@bib@innerbib\@empty
\bibitem [{\citenamefont {{Das Gupta}}\ and\ \citenamefont
  {Ghosh}(1946)}]{Gupta46}%
  \BibitemOpen
  \bibfield  {author} {\bibinfo {author} {\bibfnamefont {N.~N.}\ \bibnamefont
  {{Das Gupta}}}\ and\ \bibinfo {author} {\bibfnamefont {S.~K.}\ \bibnamefont
  {Ghosh}},\ }\bibfield  {title} {\bibinfo {title} {A report on the {W}ilson
  cloud chamber and its applications in physics},\ }\href
  {https://doi.org/10.1103/RevModPhys.18.225} {\bibfield  {journal} {\bibinfo
  {journal} {Rev. Mod. Phys.}\ }\textbf {\bibinfo {volume} {18}},\ \bibinfo
  {pages} {225} (\bibinfo {year} {1946})}\BibitemShut {NoStop}%
\bibitem [{\citenamefont {Geiger}\ and\ \citenamefont
  {Marsden}(1913)}]{Geiger13}%
  \BibitemOpen
  \bibfield  {author} {\bibinfo {author} {\bibfnamefont {H.}~\bibnamefont
  {Geiger}}\ and\ \bibinfo {author} {\bibfnamefont {E.}~\bibnamefont
  {Marsden}},\ }\bibfield  {title} {\bibinfo {title} {The laws of deflexion of
  $\alpha$ particles through large angles},\ }\href
  {https://doi.org/10.1080/14786440408634197} {\bibfield  {journal} {\bibinfo
  {journal} {Philos. Mag.}\ }\textbf {\bibinfo {volume} {25}},\ \bibinfo
  {pages} {604} (\bibinfo {year} {1913})}\BibitemShut {NoStop}%
\bibitem [{\citenamefont {Rutherford}(1911)}]{Rutherford11}%
  \BibitemOpen
  \bibfield  {author} {\bibinfo {author} {\bibfnamefont {E.}~\bibnamefont
  {Rutherford}},\ }\bibfield  {title} {\bibinfo {title} {The scattering of
  $\alpha$ and $\beta$ particles by matter and the structure of the atom},\
  }\href {https://doi.org/10.1080/14786440508637080} {\bibfield  {journal}
  {\bibinfo  {journal} {Philos. Mag.}\ }\textbf {\bibinfo {volume} {21}},\
  \bibinfo {pages} {669} (\bibinfo {year} {1911})}\BibitemShut {NoStop}%
\bibitem [{\citenamefont {M{\"u}ller}\ and\ \citenamefont
  {Bahadur}(1956)}]{Muller56}%
  \BibitemOpen
  \bibfield  {author} {\bibinfo {author} {\bibfnamefont {E.~W.}\ \bibnamefont
  {M{\"u}ller}}\ and\ \bibinfo {author} {\bibfnamefont {K.}~\bibnamefont
  {Bahadur}},\ }\bibfield  {title} {\bibinfo {title} {Field ionization of gases
  at a metal surface and the resolution of the field ion microscope},\ }\href
  {https://doi.org/10.1103/PhysRev.102.624} {\bibfield  {journal} {\bibinfo
  {journal} {Phys. Rev.}\ }\textbf {\bibinfo {volume} {102}},\ \bibinfo {pages}
  {624} (\bibinfo {year} {1956})}\BibitemShut {NoStop}%
\bibitem [{\citenamefont {Bergquist}\ \emph {et~al.}(1986)\citenamefont
  {Bergquist}, \citenamefont {Hulet}, \citenamefont {Itano},\ and\
  \citenamefont {Wineland}}]{Bergquist86}%
  \BibitemOpen
  \bibfield  {author} {\bibinfo {author} {\bibfnamefont {J.~C.}\ \bibnamefont
  {Bergquist}}, \bibinfo {author} {\bibfnamefont {R.~G.}\ \bibnamefont
  {Hulet}}, \bibinfo {author} {\bibfnamefont {W.~M.}\ \bibnamefont {Itano}},\
  and\ \bibinfo {author} {\bibfnamefont {D.~J.}\ \bibnamefont {Wineland}},\
  }\bibfield  {title} {\bibinfo {title} {Observation of quantum jumps in a
  single atom},\ }\href {https://doi.org/10.1103/PhysRevLett.57.1699}
  {\bibfield  {journal} {\bibinfo  {journal} {Phys. Rev. Lett.}\ }\textbf
  {\bibinfo {volume} {57}},\ \bibinfo {pages} {1699} (\bibinfo {year}
  {1986})}\BibitemShut {NoStop}%
\bibitem [{\citenamefont {Sauter}\ \emph {et~al.}(1986)\citenamefont {Sauter},
  \citenamefont {Neuhauser}, \citenamefont {Blatt},\ and\ \citenamefont
  {Toschek}}]{Sauter86}%
  \BibitemOpen
  \bibfield  {author} {\bibinfo {author} {\bibfnamefont {T.}~\bibnamefont
  {Sauter}}, \bibinfo {author} {\bibfnamefont {W.}~\bibnamefont {Neuhauser}},
  \bibinfo {author} {\bibfnamefont {R.}~\bibnamefont {Blatt}},\ and\ \bibinfo
  {author} {\bibfnamefont {P.~E.}\ \bibnamefont {Toschek}},\ }\bibfield
  {title} {\bibinfo {title} {Observation of quantum jumps},\ }\href
  {https://doi.org/10.1103/PhysRevLett.57.1696} {\bibfield  {journal} {\bibinfo
   {journal} {Phys. Rev. Lett.}\ }\textbf {\bibinfo {volume} {57}},\ \bibinfo
  {pages} {1696} (\bibinfo {year} {1986})}\BibitemShut {NoStop}%
\bibitem [{\citenamefont {Schellekens}\ \emph {et~al.}(2005)\citenamefont
  {Schellekens}, \citenamefont {Hoppeler}, \citenamefont {Perrin},
  \citenamefont {Gomes}, \citenamefont {Boiron}, \citenamefont {Aspect},\ and\
  \citenamefont {Westbrook}}]{Schellekens05}%
  \BibitemOpen
  \bibfield  {author} {\bibinfo {author} {\bibfnamefont {M.}~\bibnamefont
  {Schellekens}}, \bibinfo {author} {\bibfnamefont {R.}~\bibnamefont
  {Hoppeler}}, \bibinfo {author} {\bibfnamefont {A.}~\bibnamefont {Perrin}},
  \bibinfo {author} {\bibfnamefont {J.~V.}\ \bibnamefont {Gomes}}, \bibinfo
  {author} {\bibfnamefont {D.}~\bibnamefont {Boiron}}, \bibinfo {author}
  {\bibfnamefont {A.}~\bibnamefont {Aspect}},\ and\ \bibinfo {author}
  {\bibfnamefont {C.~I.}\ \bibnamefont {Westbrook}},\ }\bibfield  {title}
  {\bibinfo {title} {{H}anbury {B}rown {T}wiss effect for ultracold quantum
  gases},\ }\href {https://doi.org/10.1126/science.1118024} {\bibfield
  {journal} {\bibinfo  {journal} {Science}\ }\textbf {\bibinfo {volume}
  {310}},\ \bibinfo {pages} {648} (\bibinfo {year} {2005})}\BibitemShut
  {NoStop}%
\bibitem [{\citenamefont {Jeltes}\ \emph {et~al.}(2007)\citenamefont {Jeltes},
  \citenamefont {McNamara}, \citenamefont {Hogervorst}, \citenamefont {Vassen},
  \citenamefont {Krachmalnicoff}, \citenamefont {Schellekens}, \citenamefont
  {Perrin}, \citenamefont {Chang}, \citenamefont {Boiron}, \citenamefont
  {Aspect},\ and\ \citenamefont {Westbrook}}]{Jeltes07}%
  \BibitemOpen
  \bibfield  {author} {\bibinfo {author} {\bibfnamefont {T.}~\bibnamefont
  {Jeltes}}, \bibinfo {author} {\bibfnamefont {J.~M.}\ \bibnamefont
  {McNamara}}, \bibinfo {author} {\bibfnamefont {W.}~\bibnamefont
  {Hogervorst}}, \bibinfo {author} {\bibfnamefont {W.}~\bibnamefont {Vassen}},
  \bibinfo {author} {\bibfnamefont {V.}~\bibnamefont {Krachmalnicoff}},
  \bibinfo {author} {\bibfnamefont {M.}~\bibnamefont {Schellekens}}, \bibinfo
  {author} {\bibfnamefont {A.}~\bibnamefont {Perrin}}, \bibinfo {author}
  {\bibfnamefont {H.}~\bibnamefont {Chang}}, \bibinfo {author} {\bibfnamefont
  {D.}~\bibnamefont {Boiron}}, \bibinfo {author} {\bibfnamefont
  {A.}~\bibnamefont {Aspect}},\ and\ \bibinfo {author} {\bibfnamefont {C.~I.}\
  \bibnamefont {Westbrook}},\ }\bibfield  {title} {\bibinfo {title} {Comparison
  of the {H}anbury {B}rown-{T}wiss effect for bosons and fermions},\ }\href
  {https://doi.org/10.1038/nature05513} {\bibfield  {journal} {\bibinfo
  {journal} {Nature}\ }\textbf {\bibinfo {volume} {445}},\ \bibinfo {pages}
  {402} (\bibinfo {year} {2007})}\BibitemShut {NoStop}%
\bibitem [{\citenamefont {Gericke}\ \emph {et~al.}(2008)\citenamefont
  {Gericke}, \citenamefont {W{\"u}rtz}, \citenamefont {Reitz}, \citenamefont
  {Langen},\ and\ \citenamefont {Ott}}]{Gericke08}%
  \BibitemOpen
  \bibfield  {author} {\bibinfo {author} {\bibfnamefont {T.}~\bibnamefont
  {Gericke}}, \bibinfo {author} {\bibfnamefont {P.}~\bibnamefont {W{\"u}rtz}},
  \bibinfo {author} {\bibfnamefont {D.}~\bibnamefont {Reitz}}, \bibinfo
  {author} {\bibfnamefont {T.}~\bibnamefont {Langen}},\ and\ \bibinfo {author}
  {\bibfnamefont {H.}~\bibnamefont {Ott}},\ }\bibfield  {title} {\bibinfo
  {title} {High-resolution scanning electron microscopy of an ultracold quantum
  gas},\ }\href {https://doi.org/10.1038/nphys1102} {\bibfield  {journal}
  {\bibinfo  {journal} {Nat. Phys.}\ }\textbf {\bibinfo {volume} {4}},\
  \bibinfo {pages} {949} (\bibinfo {year} {2008})}\BibitemShut {NoStop}%
\bibitem [{\citenamefont {Ott}(2016)}]{Ott16}%
  \BibitemOpen
  \bibfield  {author} {\bibinfo {author} {\bibfnamefont {H.}~\bibnamefont
  {Ott}},\ }\bibfield  {title} {\bibinfo {title} {Single atom detection in
  ultracold quantum gases: a review of current progress},\ }\href
  {https://doi.org/10.1088/0034-4885/79/5/054401} {\bibfield  {journal}
  {\bibinfo  {journal} {Rep. Prog. Phys.}\ }\textbf {\bibinfo {volume} {79}},\
  \bibinfo {pages} {054401} (\bibinfo {year} {2016})}\BibitemShut {NoStop}%
\bibitem [{\citenamefont {Bakr}\ \emph {et~al.}(2009)\citenamefont {Bakr},
  \citenamefont {Gillen}, \citenamefont {Peng}, \citenamefont {F{\"o}lling},\
  and\ \citenamefont {Greiner}}]{Bakr09}%
  \BibitemOpen
  \bibfield  {author} {\bibinfo {author} {\bibfnamefont {W.~S.}\ \bibnamefont
  {Bakr}}, \bibinfo {author} {\bibfnamefont {J.~I.}\ \bibnamefont {Gillen}},
  \bibinfo {author} {\bibfnamefont {A.}~\bibnamefont {Peng}}, \bibinfo {author}
  {\bibfnamefont {S.}~\bibnamefont {F{\"o}lling}},\ and\ \bibinfo {author}
  {\bibfnamefont {M.}~\bibnamefont {Greiner}},\ }\bibfield  {title} {\bibinfo
  {title} {A quantum gas microscope for detecting single atoms in a
  {H}ubbard-regime optical lattice},\ }\href
  {https://doi.org/10.1038/nature08482} {\bibfield  {journal} {\bibinfo
  {journal} {Nature}\ }\textbf {\bibinfo {volume} {462}},\ \bibinfo {pages}
  {74} (\bibinfo {year} {2009})}\BibitemShut {NoStop}%
\bibitem [{\citenamefont {Gross}\ and\ \citenamefont {Bloch}(2017)}]{Gross17}%
  \BibitemOpen
  \bibfield  {author} {\bibinfo {author} {\bibfnamefont {C.}~\bibnamefont
  {Gross}}\ and\ \bibinfo {author} {\bibfnamefont {I.}~\bibnamefont {Bloch}},\
  }\bibfield  {title} {\bibinfo {title} {Quantum simulations with ultracold
  atoms in optical lattices},\ }\href {https://doi.org/10.1126/science.aal3837}
  {\bibfield  {journal} {\bibinfo  {journal} {Science}\ }\textbf {\bibinfo
  {volume} {357}},\ \bibinfo {pages} {995} (\bibinfo {year}
  {2017})}\BibitemShut {NoStop}%
\bibitem [{\citenamefont {Bakr}\ \emph {et~al.}(2010)\citenamefont {Bakr},
  \citenamefont {Peng}, \citenamefont {Tai}, \citenamefont {Ma}, \citenamefont
  {Simon}, \citenamefont {Gillen}, \citenamefont {F{\"o}lling}, \citenamefont
  {Pollet},\ and\ \citenamefont {Greiner}}]{Bakr10}%
  \BibitemOpen
  \bibfield  {author} {\bibinfo {author} {\bibfnamefont {W.~S.}\ \bibnamefont
  {Bakr}}, \bibinfo {author} {\bibfnamefont {A.}~\bibnamefont {Peng}}, \bibinfo
  {author} {\bibfnamefont {M.~E.}\ \bibnamefont {Tai}}, \bibinfo {author}
  {\bibfnamefont {R.}~\bibnamefont {Ma}}, \bibinfo {author} {\bibfnamefont
  {J.}~\bibnamefont {Simon}}, \bibinfo {author} {\bibfnamefont {J.~I.}\
  \bibnamefont {Gillen}}, \bibinfo {author} {\bibfnamefont {S.}~\bibnamefont
  {F{\"o}lling}}, \bibinfo {author} {\bibfnamefont {L.}~\bibnamefont
  {Pollet}},\ and\ \bibinfo {author} {\bibfnamefont {M.}~\bibnamefont
  {Greiner}},\ }\bibfield  {title} {\bibinfo {title} {Probing the
  superfluid--to--{M}ott insulator transition at the single-atom level},\
  }\href {https://doi.org/10.1126/science.1192368} {\bibfield  {journal}
  {\bibinfo  {journal} {Science}\ }\textbf {\bibinfo {volume} {329}},\ \bibinfo
  {pages} {547} (\bibinfo {year} {2010})}\BibitemShut {NoStop}%
\bibitem [{\citenamefont {Sherson}\ \emph {et~al.}(2010)\citenamefont
  {Sherson}, \citenamefont {Weitenberg}, \citenamefont {Endres}, \citenamefont
  {Cheneau}, \citenamefont {Bloch},\ and\ \citenamefont {Kuhr}}]{Sherson10}%
  \BibitemOpen
  \bibfield  {author} {\bibinfo {author} {\bibfnamefont {J.~F.}\ \bibnamefont
  {Sherson}}, \bibinfo {author} {\bibfnamefont {C.}~\bibnamefont {Weitenberg}},
  \bibinfo {author} {\bibfnamefont {M.}~\bibnamefont {Endres}}, \bibinfo
  {author} {\bibfnamefont {M.}~\bibnamefont {Cheneau}}, \bibinfo {author}
  {\bibfnamefont {I.}~\bibnamefont {Bloch}},\ and\ \bibinfo {author}
  {\bibfnamefont {S.}~\bibnamefont {Kuhr}},\ }\bibfield  {title} {\bibinfo
  {title} {Single-atom-resolved fluorescence imaging of an atomic {M}ott
  insulator},\ }\href {https://doi.org/10.1038/nature09378} {\bibfield
  {journal} {\bibinfo  {journal} {Nature}\ }\textbf {\bibinfo {volume} {467}},\
  \bibinfo {pages} {68} (\bibinfo {year} {2010})}\BibitemShut {NoStop}%
\bibitem [{\citenamefont {Mazurenko}\ \emph {et~al.}(2017)\citenamefont
  {Mazurenko}, \citenamefont {Chiu}, \citenamefont {Ji}, \citenamefont
  {Parsons}, \citenamefont {Kan{\'a}sz-Nagy}, \citenamefont {Schmidt},
  \citenamefont {Grusdt}, \citenamefont {Demler}, \citenamefont {Greif},\ and\
  \citenamefont {Greiner}}]{Mazurenko17}%
  \BibitemOpen
  \bibfield  {author} {\bibinfo {author} {\bibfnamefont {A.}~\bibnamefont
  {Mazurenko}}, \bibinfo {author} {\bibfnamefont {C.~S.}\ \bibnamefont {Chiu}},
  \bibinfo {author} {\bibfnamefont {G.}~\bibnamefont {Ji}}, \bibinfo {author}
  {\bibfnamefont {M.~F.}\ \bibnamefont {Parsons}}, \bibinfo {author}
  {\bibfnamefont {M.}~\bibnamefont {Kan{\'a}sz-Nagy}}, \bibinfo {author}
  {\bibfnamefont {R.}~\bibnamefont {Schmidt}}, \bibinfo {author} {\bibfnamefont
  {F.}~\bibnamefont {Grusdt}}, \bibinfo {author} {\bibfnamefont
  {E.}~\bibnamefont {Demler}}, \bibinfo {author} {\bibfnamefont
  {D.}~\bibnamefont {Greif}},\ and\ \bibinfo {author} {\bibfnamefont
  {M.}~\bibnamefont {Greiner}},\ }\bibfield  {title} {\bibinfo {title} {A
  cold-atom {F}ermi--{H}ubbard antiferromagnet},\ }\href
  {https://doi.org/10.1038/nature22362} {\bibfield  {journal} {\bibinfo
  {journal} {Nature}\ }\textbf {\bibinfo {volume} {545}},\ \bibinfo {pages}
  {462} (\bibinfo {year} {2017})}\BibitemShut {NoStop}%
\bibitem [{\citenamefont {Nichols}\ \emph {et~al.}(2019)\citenamefont
  {Nichols}, \citenamefont {Cheuk}, \citenamefont {Okan}, \citenamefont
  {Hartke}, \citenamefont {Mendez}, \citenamefont {Senthil}, \citenamefont
  {Khatami}, \citenamefont {Zhang},\ and\ \citenamefont
  {Zwierlein}}]{Nichols19}%
  \BibitemOpen
  \bibfield  {author} {\bibinfo {author} {\bibfnamefont {M.~A.}\ \bibnamefont
  {Nichols}}, \bibinfo {author} {\bibfnamefont {L.~W.}\ \bibnamefont {Cheuk}},
  \bibinfo {author} {\bibfnamefont {M.}~\bibnamefont {Okan}}, \bibinfo {author}
  {\bibfnamefont {T.~R.}\ \bibnamefont {Hartke}}, \bibinfo {author}
  {\bibfnamefont {E.}~\bibnamefont {Mendez}}, \bibinfo {author} {\bibfnamefont
  {T.}~\bibnamefont {Senthil}}, \bibinfo {author} {\bibfnamefont
  {E.}~\bibnamefont {Khatami}}, \bibinfo {author} {\bibfnamefont
  {H.}~\bibnamefont {Zhang}},\ and\ \bibinfo {author} {\bibfnamefont {M.~W.}\
  \bibnamefont {Zwierlein}},\ }\bibfield  {title} {\bibinfo {title} {Spin
  transport in a {M}ott insulator of ultracold fermions},\ }\href
  {https://doi.org/10.1126/science.aat4387} {\bibfield  {journal} {\bibinfo
  {journal} {Science}\ }\textbf {\bibinfo {volume} {363}},\ \bibinfo {pages}
  {383} (\bibinfo {year} {2019})}\BibitemShut {NoStop}%
\bibitem [{\citenamefont {McDonald}\ \emph {et~al.}(2019)\citenamefont
  {McDonald}, \citenamefont {Trisnadi}, \citenamefont {Yao},\ and\
  \citenamefont {Chin}}]{McDonald19}%
  \BibitemOpen
  \bibfield  {author} {\bibinfo {author} {\bibfnamefont {M.}~\bibnamefont
  {McDonald}}, \bibinfo {author} {\bibfnamefont {J.}~\bibnamefont {Trisnadi}},
  \bibinfo {author} {\bibfnamefont {K.-X.}\ \bibnamefont {Yao}},\ and\ \bibinfo
  {author} {\bibfnamefont {C.}~\bibnamefont {Chin}},\ }\bibfield  {title}
  {\bibinfo {title} {Superresolution microscopy of cold atoms in an optical
  lattice},\ }\href {https://doi.org/10.1103/PhysRevX.9.021001} {\bibfield
  {journal} {\bibinfo  {journal} {Phys. Rev. X}\ }\textbf {\bibinfo {volume}
  {9}},\ \bibinfo {pages} {021001} (\bibinfo {year} {2019})}\BibitemShut
  {NoStop}%
\bibitem [{\citenamefont {Subhankar}\ \emph {et~al.}(2019)\citenamefont
  {Subhankar}, \citenamefont {Wang}, \citenamefont {Tsui}, \citenamefont
  {Rolston},\ and\ \citenamefont {Porto}}]{Subhankar19}%
  \BibitemOpen
  \bibfield  {author} {\bibinfo {author} {\bibfnamefont {S.}~\bibnamefont
  {Subhankar}}, \bibinfo {author} {\bibfnamefont {Y.}~\bibnamefont {Wang}},
  \bibinfo {author} {\bibfnamefont {T.-C.}\ \bibnamefont {Tsui}}, \bibinfo
  {author} {\bibfnamefont {S.~L.}\ \bibnamefont {Rolston}},\ and\ \bibinfo
  {author} {\bibfnamefont {J.~V.}\ \bibnamefont {Porto}},\ }\bibfield  {title}
  {\bibinfo {title} {Nanoscale atomic density microscopy},\ }\href
  {https://doi.org/10.1103/PhysRevX.9.021002} {\bibfield  {journal} {\bibinfo
  {journal} {Phys. Rev. X}\ }\textbf {\bibinfo {volume} {9}},\ \bibinfo {pages}
  {021002} (\bibinfo {year} {2019})}\BibitemShut {NoStop}%
\bibitem [{\citenamefont {Nguyen}\ \emph {et~al.}(2004)\citenamefont {Nguyen},
  \citenamefont {Fléchard}, \citenamefont {Brédy}, \citenamefont {Camp},\
  and\ \citenamefont {DePaola}}]{Nguyen04}%
  \BibitemOpen
  \bibfield  {author} {\bibinfo {author} {\bibfnamefont {H.}~\bibnamefont
  {Nguyen}}, \bibinfo {author} {\bibfnamefont {X.}~\bibnamefont {Fléchard}},
  \bibinfo {author} {\bibfnamefont {R.}~\bibnamefont {Brédy}}, \bibinfo
  {author} {\bibfnamefont {H.~A.}\ \bibnamefont {Camp}},\ and\ \bibinfo
  {author} {\bibfnamefont {B.~D.}\ \bibnamefont {DePaola}},\ }\bibfield
  {title} {\bibinfo {title} {Recoil ion momentum spectroscopy using
  magneto-optically trapped atoms},\ }\href {https://doi.org/10.1063/1.1775310}
  {\bibfield  {journal} {\bibinfo  {journal} {Rev. Sci. Instrum.}\ }\textbf
  {\bibinfo {volume} {75}},\ \bibinfo {pages} {2638} (\bibinfo {year}
  {2004})}\BibitemShut {NoStop}%
\bibitem [{\citenamefont {DePaola}\ \emph {et~al.}(2008)\citenamefont
  {DePaola}, \citenamefont {Morgenstern},\ and\ \citenamefont
  {Andersen}}]{DePaola08}%
  \BibitemOpen
  \bibfield  {author} {\bibinfo {author} {\bibfnamefont {B.~D.}\ \bibnamefont
  {DePaola}}, \bibinfo {author} {\bibfnamefont {R.}~\bibnamefont
  {Morgenstern}},\ and\ \bibinfo {author} {\bibfnamefont {N.}~\bibnamefont
  {Andersen}},\ }\bibfield  {title} {\bibinfo {title} {{MOTRIMS}:
  {M}agneto–optical trap recoil ion momentum spectroscopy},\ }\href
  {https://doi.org/https://doi.org/10.1016/S1049-250X(07)55003-2} {\bibfield
  {journal} {\bibinfo  {journal} {Adv. At. Mol. Opt. Phys.}\ }\textbf {\bibinfo
  {volume} {55}},\ \bibinfo {pages} {139 } (\bibinfo {year}
  {2008})}\BibitemShut {NoStop}%
\bibitem [{\citenamefont {Stecker}\ \emph {et~al.}(2017)\citenamefont
  {Stecker}, \citenamefont {Schefzyk}, \citenamefont {Fort{\'a}gh},\ and\
  \citenamefont {G{\"u}nther}}]{Stecker17}%
  \BibitemOpen
  \bibfield  {author} {\bibinfo {author} {\bibfnamefont {M.}~\bibnamefont
  {Stecker}}, \bibinfo {author} {\bibfnamefont {H.}~\bibnamefont {Schefzyk}},
  \bibinfo {author} {\bibfnamefont {J.}~\bibnamefont {Fort{\'a}gh}},\ and\
  \bibinfo {author} {\bibfnamefont {A.}~\bibnamefont {G{\"u}nther}},\
  }\bibfield  {title} {\bibinfo {title} {A high resolution ion microscope for
  cold atoms},\ }\href {https://doi.org/10.1088/1367-2630/aa6741} {\bibfield
  {journal} {\bibinfo  {journal} {New J. Phys.}\ }\textbf {\bibinfo {volume}
  {19}},\ \bibinfo {pages} {043020} (\bibinfo {year} {2017})}\BibitemShut
  {NoStop}%
\bibitem [{\citenamefont {van Bijnen}\ \emph {et~al.}(2015)\citenamefont {van
  Bijnen}, \citenamefont {Ravensbergen}, \citenamefont {Bakker}, \citenamefont
  {Dijk}, \citenamefont {Kokkelmans},\ and\ \citenamefont
  {Vredenbregt}}]{Bijnen15}%
  \BibitemOpen
  \bibfield  {author} {\bibinfo {author} {\bibfnamefont {R.~M.~W.}\
  \bibnamefont {van Bijnen}}, \bibinfo {author} {\bibfnamefont
  {C.}~\bibnamefont {Ravensbergen}}, \bibinfo {author} {\bibfnamefont {D.~J.}\
  \bibnamefont {Bakker}}, \bibinfo {author} {\bibfnamefont {G.~J.}\
  \bibnamefont {Dijk}}, \bibinfo {author} {\bibfnamefont {S.~J. J. M.~F.}\
  \bibnamefont {Kokkelmans}},\ and\ \bibinfo {author} {\bibfnamefont
  {E.~J.~D.}\ \bibnamefont {Vredenbregt}},\ }\bibfield  {title} {\bibinfo
  {title} {Patterned {R}ydberg excitation and ionization with a spatial light
  modulator},\ }\href {https://doi.org/10.1088/1367-2630/17/2/023045}
  {\bibfield  {journal} {\bibinfo  {journal} {New J. Phys.}\ }\textbf {\bibinfo
  {volume} {17}},\ \bibinfo {pages} {023045} (\bibinfo {year}
  {2015})}\BibitemShut {NoStop}%
\bibitem [{\citenamefont {Schwarzkopf}\ \emph {et~al.}(2011)\citenamefont
  {Schwarzkopf}, \citenamefont {Sapiro},\ and\ \citenamefont
  {Raithel}}]{Schwarzkopf11}%
  \BibitemOpen
  \bibfield  {author} {\bibinfo {author} {\bibfnamefont {A.}~\bibnamefont
  {Schwarzkopf}}, \bibinfo {author} {\bibfnamefont {R.~E.}\ \bibnamefont
  {Sapiro}},\ and\ \bibinfo {author} {\bibfnamefont {G.}~\bibnamefont
  {Raithel}},\ }\bibfield  {title} {\bibinfo {title} {Imaging spatial
  correlations of {R}ydberg excitations in cold atom clouds},\ }\href
  {https://doi.org/10.1103/PhysRevLett.107.103001} {\bibfield  {journal}
  {\bibinfo  {journal} {Phys. Rev. Lett.}\ }\textbf {\bibinfo {volume} {107}},\
  \bibinfo {pages} {103001} (\bibinfo {year} {2011})}\BibitemShut {NoStop}%
\bibitem [{\citenamefont {Schwarzkopf}\ \emph {et~al.}(2013)\citenamefont
  {Schwarzkopf}, \citenamefont {Anderson}, \citenamefont {Thaicharoen},\ and\
  \citenamefont {Raithel}}]{Schwarzkopf13}%
  \BibitemOpen
  \bibfield  {author} {\bibinfo {author} {\bibfnamefont {A.}~\bibnamefont
  {Schwarzkopf}}, \bibinfo {author} {\bibfnamefont {D.~A.}\ \bibnamefont
  {Anderson}}, \bibinfo {author} {\bibfnamefont {N.}~\bibnamefont
  {Thaicharoen}},\ and\ \bibinfo {author} {\bibfnamefont {G.}~\bibnamefont
  {Raithel}},\ }\bibfield  {title} {\bibinfo {title} {Spatial correlations
  between {R}ydberg atoms in an optical dipole trap},\ }\href
  {https://doi.org/10.1103/PhysRevA.88.061406} {\bibfield  {journal} {\bibinfo
  {journal} {Phys. Rev. A}\ }\textbf {\bibinfo {volume} {88}},\ \bibinfo
  {pages} {061406} (\bibinfo {year} {2013})}\BibitemShut {NoStop}%
\bibitem [{\citenamefont {Fahey}\ \emph {et~al.}(2015)\citenamefont {Fahey},
  \citenamefont {Carroll},\ and\ \citenamefont {Noel}}]{Fahey15}%
  \BibitemOpen
  \bibfield  {author} {\bibinfo {author} {\bibfnamefont {D.~P.}\ \bibnamefont
  {Fahey}}, \bibinfo {author} {\bibfnamefont {T.~J.}\ \bibnamefont {Carroll}},\
  and\ \bibinfo {author} {\bibfnamefont {M.~W.}\ \bibnamefont {Noel}},\
  }\bibfield  {title} {\bibinfo {title} {Imaging the dipole-dipole energy
  exchange between ultracold rubidium {R}ydberg atoms},\ }\href
  {https://doi.org/10.1103/PhysRevA.91.062702} {\bibfield  {journal} {\bibinfo
  {journal} {Phys. Rev. A}\ }\textbf {\bibinfo {volume} {91}},\ \bibinfo
  {pages} {062702} (\bibinfo {year} {2015})}\BibitemShut {NoStop}%
\bibitem [{\citenamefont {Stecker}\ \emph {et~al.}(2019)\citenamefont
  {Stecker}, \citenamefont {Nold}, \citenamefont {Steinert}, \citenamefont
  {Grimmel}, \citenamefont {Petrosyan}, \citenamefont {Fortágh},\ and\
  \citenamefont {Günther}}]{stecker19}%
  \BibitemOpen
  \bibfield  {author} {\bibinfo {author} {\bibfnamefont {M.}~\bibnamefont
  {Stecker}}, \bibinfo {author} {\bibfnamefont {R.}~\bibnamefont {Nold}},
  \bibinfo {author} {\bibfnamefont {L.-M.}\ \bibnamefont {Steinert}}, \bibinfo
  {author} {\bibfnamefont {J.}~\bibnamefont {Grimmel}}, \bibinfo {author}
  {\bibfnamefont {D.}~\bibnamefont {Petrosyan}}, \bibinfo {author}
  {\bibfnamefont {J.}~\bibnamefont {Fortágh}},\ and\ \bibinfo {author}
  {\bibfnamefont {A.}~\bibnamefont {Günther}},\ }\href@noop {} {\bibinfo
  {title} {Controlling the dipole blockade and ionization rate of {R}ydberg
  atoms in strong electric fields}} (\bibinfo {year} {2019}),\ \Eprint
  {https://arxiv.org/abs/1905.08221} {arXiv:1905.08221} \BibitemShut {NoStop}%
\bibitem [{\citenamefont {Wiza}(1979)}]{Wiza79}%
  \BibitemOpen
  \bibfield  {author} {\bibinfo {author} {\bibfnamefont {J.~L.}\ \bibnamefont
  {Wiza}},\ }\bibfield  {title} {\bibinfo {title} {Microchannel plate
  detectors},\ }\href
  {https://doi.org/https://doi.org/10.1016/0029-554X(79)90734-1} {\bibfield
  {journal} {\bibinfo  {journal} {Nucl. Instrum. Methods}\ }\textbf {\bibinfo
  {volume} {162}},\ \bibinfo {pages} {587 } (\bibinfo {year}
  {1979})}\BibitemShut {NoStop}%
\bibitem [{\citenamefont {Jagutzki}\ \emph {et~al.}(2002)\citenamefont
  {Jagutzki}, \citenamefont {Mergel}, \citenamefont {Ullmann-Pfleger},
  \citenamefont {Spielberger}, \citenamefont {Spillmann}, \citenamefont
  {Dörner},\ and\ \citenamefont {Schmidt-Böcking}}]{Jagutzki02}%
  \BibitemOpen
  \bibfield  {author} {\bibinfo {author} {\bibfnamefont {O.}~\bibnamefont
  {Jagutzki}}, \bibinfo {author} {\bibfnamefont {V.}~\bibnamefont {Mergel}},
  \bibinfo {author} {\bibfnamefont {K.}~\bibnamefont {Ullmann-Pfleger}},
  \bibinfo {author} {\bibfnamefont {L.}~\bibnamefont {Spielberger}}, \bibinfo
  {author} {\bibfnamefont {U.}~\bibnamefont {Spillmann}}, \bibinfo {author}
  {\bibfnamefont {R.}~\bibnamefont {Dörner}},\ and\ \bibinfo {author}
  {\bibfnamefont {H.}~\bibnamefont {Schmidt-Böcking}},\ }\bibfield  {title}
  {\bibinfo {title} {A broad-application microchannel-plate detector system for
  advanced particle or photon detection tasks: large area imaging, precise
  multi-hit timing information and high detection rate},\ }\href
  {https://doi.org/https://doi.org/10.1016/S0168-9002(01)01839-3} {\bibfield
  {journal} {\bibinfo  {journal} {Nucl. Instrum. Methods Phys. Res.}\ }\textbf
  {\bibinfo {volume} {477}},\ \bibinfo {pages} {244 } (\bibinfo {year}
  {2002})}\BibitemShut {NoStop}%
\bibitem [{\citenamefont {Schmid}(2019)}]{Schmid19}%
  \BibitemOpen
  \bibfield  {author} {\bibinfo {author} {\bibfnamefont {T.}~\bibnamefont
  {Schmid}},\ }\emph {\bibinfo {title} {{R}ydberg molecules for ultracold
  ion-atom scattering}},\ \href@noop {} {Ph.D. thesis},\ \bibinfo  {school}
  {Universit{\"a}t Stuttgart} (\bibinfo {year} {2019})\BibitemShut {NoStop}%
\bibitem [{\citenamefont {Engel}\ \emph {et~al.}(2018)\citenamefont {Engel},
  \citenamefont {Dieterle}, \citenamefont {Schmid}, \citenamefont {Tomschitz},
  \citenamefont {Veit}, \citenamefont {Zuber}, \citenamefont {L\"ow},
  \citenamefont {Pfau},\ and\ \citenamefont {Meinert}}]{Engel18}%
  \BibitemOpen
  \bibfield  {author} {\bibinfo {author} {\bibfnamefont {F.}~\bibnamefont
  {Engel}}, \bibinfo {author} {\bibfnamefont {T.}~\bibnamefont {Dieterle}},
  \bibinfo {author} {\bibfnamefont {T.}~\bibnamefont {Schmid}}, \bibinfo
  {author} {\bibfnamefont {C.}~\bibnamefont {Tomschitz}}, \bibinfo {author}
  {\bibfnamefont {C.}~\bibnamefont {Veit}}, \bibinfo {author} {\bibfnamefont
  {N.}~\bibnamefont {Zuber}}, \bibinfo {author} {\bibfnamefont
  {R.}~\bibnamefont {L\"ow}}, \bibinfo {author} {\bibfnamefont
  {T.}~\bibnamefont {Pfau}},\ and\ \bibinfo {author} {\bibfnamefont
  {F.}~\bibnamefont {Meinert}},\ }\bibfield  {title} {\bibinfo {title}
  {Observation of {R}ydberg blockade induced by a single ion},\ }\href
  {https://doi.org/10.1103/PhysRevLett.121.193401} {\bibfield  {journal}
  {\bibinfo  {journal} {Phys. Rev. Lett.}\ }\textbf {\bibinfo {volume} {121}},\
  \bibinfo {pages} {193401} (\bibinfo {year} {2018})}\BibitemShut {NoStop}%
\bibitem [{\citenamefont {Tomza}\ \emph {et~al.}(2019)\citenamefont {Tomza},
  \citenamefont {Jachymski}, \citenamefont {Gerritsma}, \citenamefont
  {Negretti}, \citenamefont {Calarco}, \citenamefont {Idziaszek},\ and\
  \citenamefont {Julienne}}]{Tomza19}%
  \BibitemOpen
  \bibfield  {author} {\bibinfo {author} {\bibfnamefont {M.}~\bibnamefont
  {Tomza}}, \bibinfo {author} {\bibfnamefont {K.}~\bibnamefont {Jachymski}},
  \bibinfo {author} {\bibfnamefont {R.}~\bibnamefont {Gerritsma}}, \bibinfo
  {author} {\bibfnamefont {A.}~\bibnamefont {Negretti}}, \bibinfo {author}
  {\bibfnamefont {T.}~\bibnamefont {Calarco}}, \bibinfo {author} {\bibfnamefont
  {Z.}~\bibnamefont {Idziaszek}},\ and\ \bibinfo {author} {\bibfnamefont
  {P.~S.}\ \bibnamefont {Julienne}},\ }\bibfield  {title} {\bibinfo {title}
  {Cold hybrid ion-atom systems},\ }\href
  {https://doi.org/10.1103/RevModPhys.91.035001} {\bibfield  {journal}
  {\bibinfo  {journal} {Rev. Mod. Phys.}\ }\textbf {\bibinfo {volume} {91}},\
  \bibinfo {pages} {035001} (\bibinfo {year} {2019})}\BibitemShut {NoStop}%
\bibitem [{\citenamefont {Dieterle}\ \emph {et~al.}(2020)\citenamefont
  {Dieterle}, \citenamefont {Berngruber}, \citenamefont {H{\"o}lzl},
  \citenamefont {L{\"o}w}, \citenamefont {Jachymski}, \citenamefont {Pfau},\
  and\ \citenamefont {Meinert}}]{dieterle20}%
  \BibitemOpen
  \bibfield  {author} {\bibinfo {author} {\bibfnamefont {T.}~\bibnamefont
  {Dieterle}}, \bibinfo {author} {\bibfnamefont {M.}~\bibnamefont
  {Berngruber}}, \bibinfo {author} {\bibfnamefont {C.}~\bibnamefont
  {H{\"o}lzl}}, \bibinfo {author} {\bibfnamefont {R.}~\bibnamefont {L{\"o}w}},
  \bibinfo {author} {\bibfnamefont {K.}~\bibnamefont {Jachymski}}, \bibinfo
  {author} {\bibfnamefont {T.}~\bibnamefont {Pfau}},\ and\ \bibinfo {author}
  {\bibfnamefont {F.}~\bibnamefont {Meinert}},\ }\href@noop {} {\bibinfo
  {title} {Transport of a single cold ion immersed in a {B}ose-{E}instein
  condensate}} (\bibinfo {year} {2020}),\ \Eprint
  {https://arxiv.org/abs/2007.00309} {arXiv:2007.00309} \BibitemShut {NoStop}%
\bibitem [{\citenamefont {Astrakharchik}\ \emph {et~al.}(2020)\citenamefont
  {Astrakharchik}, \citenamefont {{Pe\~{n}a Ardila}}, \citenamefont {Schmidt},
  \citenamefont {Jachymski},\ and\ \citenamefont {Negretti}}]{astrakharchik20}%
  \BibitemOpen
  \bibfield  {author} {\bibinfo {author} {\bibfnamefont {G.~E.}\ \bibnamefont
  {Astrakharchik}}, \bibinfo {author} {\bibfnamefont {L.~A.}\ \bibnamefont
  {{Pe\~{n}a Ardila}}}, \bibinfo {author} {\bibfnamefont {R.}~\bibnamefont
  {Schmidt}}, \bibinfo {author} {\bibfnamefont {K.}~\bibnamefont {Jachymski}},\
  and\ \bibinfo {author} {\bibfnamefont {A.}~\bibnamefont {Negretti}},\
  }\href@noop {} {\bibinfo {title} {Ionic polaron in a {B}ose-{E}instein
  condensate}} (\bibinfo {year} {2020}),\ \Eprint
  {https://arxiv.org/abs/2005.12033} {arXiv:2005.12033} \BibitemShut {NoStop}%
\bibitem [{\citenamefont {Feldker}\ \emph {et~al.}(2020)\citenamefont
  {Feldker}, \citenamefont {F{\"u}rst}, \citenamefont {Hirzler}, \citenamefont
  {Ewald}, \citenamefont {Mazzanti}, \citenamefont {Wiater}, \citenamefont
  {Tomza},\ and\ \citenamefont {Gerritsma}}]{feldker20}%
  \BibitemOpen
  \bibfield  {author} {\bibinfo {author} {\bibfnamefont {T.}~\bibnamefont
  {Feldker}}, \bibinfo {author} {\bibfnamefont {H.}~\bibnamefont {F{\"u}rst}},
  \bibinfo {author} {\bibfnamefont {H.}~\bibnamefont {Hirzler}}, \bibinfo
  {author} {\bibfnamefont {N.~V.}\ \bibnamefont {Ewald}}, \bibinfo {author}
  {\bibfnamefont {M.}~\bibnamefont {Mazzanti}}, \bibinfo {author}
  {\bibfnamefont {D.}~\bibnamefont {Wiater}}, \bibinfo {author} {\bibfnamefont
  {M.}~\bibnamefont {Tomza}},\ and\ \bibinfo {author} {\bibfnamefont
  {R.}~\bibnamefont {Gerritsma}},\ }\bibfield  {title} {\bibinfo {title}
  {Buffer gas cooling of a trapped ion to the quantum regime},\ }\href
  {https://doi.org/10.1038/s41567-019-0772-5} {\bibfield  {journal} {\bibinfo
  {journal} {Nat. Phys.}\ }\textbf {\bibinfo {volume} {16}},\ \bibinfo {pages}
  {413} (\bibinfo {year} {2020})}\BibitemShut {NoStop}%
\bibitem [{\citenamefont {Schmid}\ \emph {et~al.}(2018)\citenamefont {Schmid},
  \citenamefont {Veit}, \citenamefont {Zuber}, \citenamefont {L\"ow},
  \citenamefont {Pfau}, \citenamefont {Tarana},\ and\ \citenamefont
  {Tomza}}]{Schmid18}%
  \BibitemOpen
  \bibfield  {author} {\bibinfo {author} {\bibfnamefont {T.}~\bibnamefont
  {Schmid}}, \bibinfo {author} {\bibfnamefont {C.}~\bibnamefont {Veit}},
  \bibinfo {author} {\bibfnamefont {N.}~\bibnamefont {Zuber}}, \bibinfo
  {author} {\bibfnamefont {R.}~\bibnamefont {L\"ow}}, \bibinfo {author}
  {\bibfnamefont {T.}~\bibnamefont {Pfau}}, \bibinfo {author} {\bibfnamefont
  {M.}~\bibnamefont {Tarana}},\ and\ \bibinfo {author} {\bibfnamefont
  {M.}~\bibnamefont {Tomza}},\ }\bibfield  {title} {\bibinfo {title} {{R}ydberg
  molecules for ion-atom scattering in the ultracold regime},\ }\href
  {https://doi.org/10.1103/PhysRevLett.120.153401} {\bibfield  {journal}
  {\bibinfo  {journal} {Phys. Rev. Lett.}\ }\textbf {\bibinfo {volume} {120}},\
  \bibinfo {pages} {153401} (\bibinfo {year} {2018})}\BibitemShut {NoStop}%
\bibitem [{\citenamefont {Schau{\ss}}\ \emph {et~al.}(2012)\citenamefont
  {Schau{\ss}}, \citenamefont {Cheneau}, \citenamefont {Endres}, \citenamefont
  {Fukuhara}, \citenamefont {Hild}, \citenamefont {Omran}, \citenamefont
  {Pohl}, \citenamefont {Gross}, \citenamefont {Kuhr},\ and\ \citenamefont
  {Bloch}}]{Schauss12}%
  \BibitemOpen
  \bibfield  {author} {\bibinfo {author} {\bibfnamefont {P.}~\bibnamefont
  {Schau{\ss}}}, \bibinfo {author} {\bibfnamefont {M.}~\bibnamefont {Cheneau}},
  \bibinfo {author} {\bibfnamefont {M.}~\bibnamefont {Endres}}, \bibinfo
  {author} {\bibfnamefont {T.}~\bibnamefont {Fukuhara}}, \bibinfo {author}
  {\bibfnamefont {S.}~\bibnamefont {Hild}}, \bibinfo {author} {\bibfnamefont
  {A.}~\bibnamefont {Omran}}, \bibinfo {author} {\bibfnamefont
  {T.}~\bibnamefont {Pohl}}, \bibinfo {author} {\bibfnamefont {C.}~\bibnamefont
  {Gross}}, \bibinfo {author} {\bibfnamefont {S.}~\bibnamefont {Kuhr}},\ and\
  \bibinfo {author} {\bibfnamefont {I.}~\bibnamefont {Bloch}},\ }\bibfield
  {title} {\bibinfo {title} {Observation of spatially ordered structures in a
  two-dimensional {R}ydberg gas},\ }\href {https://doi.org/10.1038/nature11596}
  {\bibfield  {journal} {\bibinfo  {journal} {Nature}\ }\textbf {\bibinfo
  {volume} {491}},\ \bibinfo {pages} {87} (\bibinfo {year} {2012})}\BibitemShut
  {NoStop}%
\bibitem [{\citenamefont {Riechers}\ \emph {et~al.}(2017)\citenamefont
  {Riechers}, \citenamefont {Hueck}, \citenamefont {Luick}, \citenamefont
  {Lompe},\ and\ \citenamefont {Moritz}}]{Riechers17}%
  \BibitemOpen
  \bibfield  {author} {\bibinfo {author} {\bibfnamefont {K.}~\bibnamefont
  {Riechers}}, \bibinfo {author} {\bibfnamefont {K.}~\bibnamefont {Hueck}},
  \bibinfo {author} {\bibfnamefont {N.}~\bibnamefont {Luick}}, \bibinfo
  {author} {\bibfnamefont {T.}~\bibnamefont {Lompe}},\ and\ \bibinfo {author}
  {\bibfnamefont {H.}~\bibnamefont {Moritz}},\ }\bibfield  {title} {\bibinfo
  {title} {Detecting {F}riedel oscillations in ultracold {F}ermi gases},\
  }\href {https://doi.org/10.1140/epjd/e2017-80275-6} {\bibfield  {journal}
  {\bibinfo  {journal} {Eur. Phys. J. D}\ }\textbf {\bibinfo {volume} {71}},\
  \bibinfo {pages} {232} (\bibinfo {year} {2017})}\BibitemShut {NoStop}%
\bibitem [{\citenamefont {Szilagyi}(1988)}]{Szilagyi12}%
  \BibitemOpen
  \bibfield  {author} {\bibinfo {author} {\bibfnamefont {M.}~\bibnamefont
  {Szilagyi}},\ }\href@noop {} {\emph {\bibinfo {title} {Electron and ion
  optics}}}\ (\bibinfo  {publisher} {Plenum Press},\ \bibinfo {year}
  {1988})\BibitemShut {NoStop}%
\bibitem [{\citenamefont {L\"ow}\ \emph {et~al.}(2012)\citenamefont {L\"ow},
  \citenamefont {Weimer}, \citenamefont {Nipper}, \citenamefont {Balewski},
  \citenamefont {Butscher}, \citenamefont {B\"uchler},\ and\ \citenamefont
  {Pfau}}]{Loew12}%
  \BibitemOpen
  \bibfield  {author} {\bibinfo {author} {\bibfnamefont {R.}~\bibnamefont
  {L\"ow}}, \bibinfo {author} {\bibfnamefont {H.}~\bibnamefont {Weimer}},
  \bibinfo {author} {\bibfnamefont {J.}~\bibnamefont {Nipper}}, \bibinfo
  {author} {\bibfnamefont {J.~B.}\ \bibnamefont {Balewski}}, \bibinfo {author}
  {\bibfnamefont {B.}~\bibnamefont {Butscher}}, \bibinfo {author}
  {\bibfnamefont {H.~P.}\ \bibnamefont {B\"uchler}},\ and\ \bibinfo {author}
  {\bibfnamefont {T.}~\bibnamefont {Pfau}},\ }\bibfield  {title} {\bibinfo
  {title} {An experimental and theoretical guide to strongly interacting
  {R}ydberg gases},\ }\href {https://doi.org/10.1088/0953-4075/45/11/113001}
  {\bibfield  {journal} {\bibinfo  {journal} {J. Phys. B}\ }\textbf {\bibinfo
  {volume} {45}},\ \bibinfo {pages} {113001} (\bibinfo {year}
  {2012})}\BibitemShut {NoStop}%
\bibitem [{Det()}]{Detector}%
  \BibitemOpen
  \href@noop {} {}\bibinfo {note} {DLD40EP, RoentDek}\BibitemShut {NoStop}%
\bibitem [{\citenamefont {Oberheide}\ \emph {et~al.}(1997)\citenamefont
  {Oberheide}, \citenamefont {Wilhelms},\ and\ \citenamefont
  {Zimmer}}]{Oberheide97}%
  \BibitemOpen
  \bibfield  {author} {\bibinfo {author} {\bibfnamefont {J.}~\bibnamefont
  {Oberheide}}, \bibinfo {author} {\bibfnamefont {P.}~\bibnamefont
  {Wilhelms}},\ and\ \bibinfo {author} {\bibfnamefont {M.}~\bibnamefont
  {Zimmer}},\ }\bibfield  {title} {\bibinfo {title} {New results on the
  absolute ion detection efficiencies of a microchannel plate},\ }\href
  {https://doi.org/10.1088/0957-0233/8/4/001} {\bibfield  {journal} {\bibinfo
  {journal} {Meas. Sci. Technol.}\ }\textbf {\bibinfo {volume} {8}},\ \bibinfo
  {pages} {351} (\bibinfo {year} {1997})}\BibitemShut {NoStop}%
\bibitem [{\citenamefont {Fehre}\ \emph {et~al.}(2018)\citenamefont {Fehre},
  \citenamefont {Trojanowskaja}, \citenamefont {Gatzke}, \citenamefont
  {Kunitski}, \citenamefont {Trinter}, \citenamefont {Zeller}, \citenamefont
  {Schmidt}, \citenamefont {Stohner}, \citenamefont {Berger}, \citenamefont
  {Czasch}, \citenamefont {Jagutzki}, \citenamefont {Jahnke}, \citenamefont
  {Dörner},\ and\ \citenamefont {Schöffler}}]{Fehre18}%
  \BibitemOpen
  \bibfield  {author} {\bibinfo {author} {\bibfnamefont {K.}~\bibnamefont
  {Fehre}}, \bibinfo {author} {\bibfnamefont {D.}~\bibnamefont
  {Trojanowskaja}}, \bibinfo {author} {\bibfnamefont {J.}~\bibnamefont
  {Gatzke}}, \bibinfo {author} {\bibfnamefont {M.}~\bibnamefont {Kunitski}},
  \bibinfo {author} {\bibfnamefont {F.}~\bibnamefont {Trinter}}, \bibinfo
  {author} {\bibfnamefont {S.}~\bibnamefont {Zeller}}, \bibinfo {author}
  {\bibfnamefont {L.~P.~H.}\ \bibnamefont {Schmidt}}, \bibinfo {author}
  {\bibfnamefont {J.}~\bibnamefont {Stohner}}, \bibinfo {author} {\bibfnamefont
  {R.}~\bibnamefont {Berger}}, \bibinfo {author} {\bibfnamefont
  {A.}~\bibnamefont {Czasch}}, \bibinfo {author} {\bibfnamefont
  {O.}~\bibnamefont {Jagutzki}}, \bibinfo {author} {\bibfnamefont
  {T.}~\bibnamefont {Jahnke}}, \bibinfo {author} {\bibfnamefont
  {R.}~\bibnamefont {Dörner}},\ and\ \bibinfo {author} {\bibfnamefont {M.~S.}\
  \bibnamefont {Schöffler}},\ }\bibfield  {title} {\bibinfo {title} {Absolute
  ion detection efficiencies of microchannel plates and funnel microchannel
  plates for multi-coincidence detection},\ }\href
  {https://doi.org/10.1063/1.5022564} {\bibfield  {journal} {\bibinfo
  {journal} {Rev. Sci. Instrum.}\ }\textbf {\bibinfo {volume} {89}},\ \bibinfo
  {pages} {045112} (\bibinfo {year} {2018})}\BibitemShut {NoStop}%
\bibitem [{\citenamefont {Urvoy}\ \emph {et~al.}(2015)\citenamefont {Urvoy},
  \citenamefont {Ripka}, \citenamefont {Lesanovsky}, \citenamefont {Booth},
  \citenamefont {Shaffer}, \citenamefont {Pfau},\ and\ \citenamefont
  {L\"ow}}]{Urvoy15}%
  \BibitemOpen
  \bibfield  {author} {\bibinfo {author} {\bibfnamefont {A.}~\bibnamefont
  {Urvoy}}, \bibinfo {author} {\bibfnamefont {F.}~\bibnamefont {Ripka}},
  \bibinfo {author} {\bibfnamefont {I.}~\bibnamefont {Lesanovsky}}, \bibinfo
  {author} {\bibfnamefont {D.}~\bibnamefont {Booth}}, \bibinfo {author}
  {\bibfnamefont {J.~P.}\ \bibnamefont {Shaffer}}, \bibinfo {author}
  {\bibfnamefont {T.}~\bibnamefont {Pfau}},\ and\ \bibinfo {author}
  {\bibfnamefont {R.}~\bibnamefont {L\"ow}},\ }\bibfield  {title} {\bibinfo
  {title} {Strongly correlated growth of {R}ydberg aggregates in a vapor
  cell},\ }\href {https://doi.org/10.1103/PhysRevLett.114.203002} {\bibfield
  {journal} {\bibinfo  {journal} {Phys. Rev. Lett.}\ }\textbf {\bibinfo
  {volume} {114}},\ \bibinfo {pages} {203002} (\bibinfo {year}
  {2015})}\BibitemShut {NoStop}%
\bibitem [{Sim()}]{Simion}%
  \BibitemOpen
  \href@noop {} {}\bibinfo {note} {SIMION 8.1.1.32, Scientific Instrument
  Services}\BibitemShut {NoStop}%
\bibitem [{\citenamefont {Abo-Shaeer}\ \emph {et~al.}(2001)\citenamefont
  {Abo-Shaeer}, \citenamefont {Raman}, \citenamefont {Vogels},\ and\
  \citenamefont {Ketterle}}]{Abo01}%
  \BibitemOpen
  \bibfield  {author} {\bibinfo {author} {\bibfnamefont {J.~R.}\ \bibnamefont
  {Abo-Shaeer}}, \bibinfo {author} {\bibfnamefont {C.}~\bibnamefont {Raman}},
  \bibinfo {author} {\bibfnamefont {J.~M.}\ \bibnamefont {Vogels}},\ and\
  \bibinfo {author} {\bibfnamefont {W.}~\bibnamefont {Ketterle}},\ }\bibfield
  {title} {\bibinfo {title} {Observation of vortex lattices in
  {B}ose-{E}instein condensates},\ }\href
  {https://doi.org/10.1126/science.1060182} {\bibfield  {journal} {\bibinfo
  {journal} {Science}\ }\textbf {\bibinfo {volume} {292}},\ \bibinfo {pages}
  {476} (\bibinfo {year} {2001})}\BibitemShut {NoStop}%
\bibitem [{\citenamefont {Marti}\ \emph {et~al.}(2010)\citenamefont {Marti},
  \citenamefont {Olf}, \citenamefont {Vogt}, \citenamefont {\"Ottl},\ and\
  \citenamefont {Stamper-Kurn}}]{Marti10}%
  \BibitemOpen
  \bibfield  {author} {\bibinfo {author} {\bibfnamefont {G.~E.}\ \bibnamefont
  {Marti}}, \bibinfo {author} {\bibfnamefont {R.}~\bibnamefont {Olf}}, \bibinfo
  {author} {\bibfnamefont {E.}~\bibnamefont {Vogt}}, \bibinfo {author}
  {\bibfnamefont {A.}~\bibnamefont {\"Ottl}},\ and\ \bibinfo {author}
  {\bibfnamefont {D.~M.}\ \bibnamefont {Stamper-Kurn}},\ }\bibfield  {title}
  {\bibinfo {title} {Two-element {Z}eeman slower for rubidium and lithium},\
  }\href {https://doi.org/10.1103/PhysRevA.81.043424} {\bibfield  {journal}
  {\bibinfo  {journal} {Phys. Rev. A}\ }\textbf {\bibinfo {volume} {81}},\
  \bibinfo {pages} {043424} (\bibinfo {year} {2010})}\BibitemShut {NoStop}%
\bibitem [{\citenamefont {Gross}\ \emph {et~al.}(2016)\citenamefont {Gross},
  \citenamefont {Gan},\ and\ \citenamefont {Dieckmann}}]{Gross16}%
  \BibitemOpen
  \bibfield  {author} {\bibinfo {author} {\bibfnamefont {C.}~\bibnamefont
  {Gross}}, \bibinfo {author} {\bibfnamefont {H.~C.~J.}\ \bibnamefont {Gan}},\
  and\ \bibinfo {author} {\bibfnamefont {K.}~\bibnamefont {Dieckmann}},\
  }\bibfield  {title} {\bibinfo {title} {All-optical production and transport
  of a large $^{6}\mathrm{Li}$ quantum gas in a crossed optical dipole trap},\
  }\href {https://doi.org/10.1103/PhysRevA.93.053424} {\bibfield  {journal}
  {\bibinfo  {journal} {Phys. Rev. A}\ }\textbf {\bibinfo {volume} {93}},\
  \bibinfo {pages} {053424} (\bibinfo {year} {2016})}\BibitemShut {NoStop}%
\end{thebibliography}
\end{document}